\documentclass[12pt]{article}
\usepackage[normalem]{ulem}
\usepackage{xcolor}
\usepackage[english]{babel}
\usepackage[T1]{fontenc}
\usepackage[utf8]{inputenc}
\usepackage{authblk}
\usepackage{mathtools}
\usepackage{slashed}
\usepackage{amsmath,amssymb}
\usepackage{amsfonts}
\usepackage{enumitem}
\usepackage{graphicx,color,xcolor}
\usepackage{cite}
\usepackage{tabularx}
\usepackage{subcaption}
\usepackage{hyperref}
\usepackage[capitalize]{cleveref}

\hypersetup{
    colorlinks=true,
    linkcolor=blue,
    filecolor=magenta,
    urlcolor=cyan,
}
\numberwithin{equation}{section}
\definecolor{refcol}{rgb}{0.9,0.1,0.1}
\hypersetup{colorlinks=true,linkcolor=blue,citecolor=refcol,urlcolor=cyan,linktocpage}
\graphicspath{{Plots/}}

\begin{document}
\include{adefn}

\begin{titlepage}
\thispagestyle{empty}

\title{
{\Huge\bf Near-Extremal Freudenthal Duality}
}

\vfill

\author{
	{\bf Arghya Chattopadhyay$^{a,b}$}\thanks{{\tt arghya.chattopadhyay@umons.ac.be}},
	{\bf Taniya Mandal$^b$}\thanks{{\tt taniya.mandal@wits.ac.za}},
	{\bf Alessio Marrani$^{c}$}\thanks{{\tt alessio.marrani@um.es}}
	\smallskip\hfill\\      	
	\small{
        $^a${\it Service de Physique de l'Univers, Champs et Gravitation}\\
		{\it Université de Mons, 20 Place du Parc, 7000 Mons, Belgium}\\
		\hfill
 \smallskip\hfill\\
		$^b${\it National Institute of Theoretical and Computational Sciences}\\
		{\it School of Physics and Mandelstam Institute for Theoretical Physics}\\
		{\it University of the Witwatersrand, Wits, 2050, South Africa}\\
		\hfill
 \smallskip\hfill\\
	    $^c${\it Instituto de  Teorica F\'{\i}sica, Dep.to de F\'{\i}sica}\\
	    {\it Universidad de Murcia, Campus de Espinardo, E-30100, Spain}\\
			}
		}

\vfill

\date{
\vspace{1cm}
\begin{quote}
\centerline{{\bf Abstract}}
{\small
Freudenthal duality is, as of now, the unique non-linear map on electric-magnetic (e.m.) charges which is a symmetry of the Bekenstein-Hawking entropy of extremal black holes, displaying the Attractor Mechanism (possibly, up to some flat directions) in Maxwell-Einstein-scalar theories in four space-time dimensions and with non-trivial symplectic e.m. duality. In this paper, we put forward an effective approach to a consistent generalization of Freudenthal duality to \textit{near-extremal} black holes, whose entropy is obtained within a Jackiw-Teitelboim gravity upon dimensional reduction. We name such a generalization \textit{near-extremal Freudenthal duality}. Upon such a duality, two near-extremal black holes with two different (and both small) temperatures have the same entropy when their e.m. charges are related by a Freudenthal transformation. By exploiting Descartes' rule of signs as well as Sturm's Theorem, we show that our formulation of the near-extremal Freudenthal duality is analytical and unique.}
\end{quote}
}



%
\end{titlepage}
\thispagestyle{empty}\maketitle\vfill \eject

\newpage

\tableofcontents

\newpage


\section{Introduction}

Extremal black holes are objects of great interest to the string theorists,
though they concern a pretty ideal situation, as they lack temperature $T$.
In the recent past, \textit{near-extremal} black holes have also been
objects of intense study. A near-extremal black hole is a black hole which
is not far from saturating the extremality bound, namely which is not far
from having the minimal possible mass that can be compatible with the
charges, and possible angular momentum (or momenta), of the black hole
itself. In supersymmetric theories, near-extremal black holes are often
small perturbations of supersymmetric (BPS) black holes; as such,
near-extremal black holes have small Hawking temperatures and consequently
emit small amounts of Hawking radiation.

Jackiw-Teitelboim (JT) gravity, introduced in the 80's~\cite%
{Teitelboim:1983ux,Jackiw:1984je}, has recently been exploited in order to
study higher-dimensional near-extremal black holes within the
two-dimensional dilaton gravity theory obtained upon dimensional reduction~%
\cite{Maldacena:2016upp,Almheiri:2014cka}. In this framework, it has been
shown that as an $AdS_{2}$ factor emerges within the near-horizon geometry
of a near-extremal black hole, its dynamics can be effectively described by
a two-dimensional dilaton gravity, obtained by dimensional reduction. This
results in to a dramatic simplification, opening up new directions within
the study of near-extremal black holes. For instance, in recent years, a
broad variety of near-extremal black holes has been studied through JT
gravity in great detail, ranging from charged to rotational black holes~\cite%
{Nayak:2018qej,Moitra:2019bub}, for non-relativistic theories~\cite%
{Kolekar:2018sba} as well as for higher-derivative gravity~\cite%
{Banerjee:2021vjy}.

A natural question arising in this context is whether the properties of
extremal black holes still persist, and, if and how they get changed or
deformed when one (slightly) departs from extremality by considering
near-extremal black holes. In this paper, we will focus on the \textit{%
Freudenthal duality} \cite{Borsten:2009zy, Ferrara:2011gv}, by investigating
whether one can make sense at all of such a property of extremal black holes
in Maxwell-Einstein (super)gravity theories in four-space time dimensions
when considering near-extremal black holes. Freudenthal duality is an
intrinsically non-linear symmetry of the Bekenstein-Hawking entropy of
extremal black holes, which, in the case of asymptotically flat and static
solutions is also an anti-involutive map in the space of electric and
magnetic charges of the black holes themselves \cite%
{Borsten:2009zy,Ferrara:2011gv,Borsten:2012pd} (cf. discussion in Sec. \ref%
{T=0-inv}). Within its anti-involutive formulation, Freudenthal duality has
been extended to act also on the fluxes supporting flux compactification of
string theory, giving rise to gauged supergravity theories in four
dimensions \cite{Klemm:2017xxk}; further investigations have been made e.g.
in \cite{Marrani:2012uu}-\nocite%
{Galli:2012ji,Fernandez-Melgarejo:2013ksa,Mandal:2017ioi,Borsten:2018djw}%
\cite{Borsten:2019xas}. Under Freudenthal duality, dyonic charges and fluxes
(i.e., gauging parameters) undergo non-linear transformations, which
therefore cannot be regarded as electric-magnetic duality (U-duality)
transformations, because the latter is linearly realized in the framework
under consideration \cite{Marrani:2015wra,Marrani:2017ihg,Marrani:2019zsn}.
Remarkably, in extremal black holes also the attractor points, describing
the attractor configurations of scalar fields at the (unique) event horizon
of the black hole, are invariant under Freudenthal duality \cite%
{Ferrara:2011gv}.

In \cite{Chattopadhyay:2021vog}, an attempt has first been made in order to
establish how Freudenthal duality acts on \textit{near-extremal} black
holes: the outcome of such an investigation resulted in the statement that
two near-extremal black holes cannot be Freudenthal dual (i.e., cannot have
the same entropy and temperature, with electric-magnetic (e.m.) charges
related by a Freudenthal duality transformation) if their e.m. charges are
not related via the Freudenthal duality transformation defined by the
corresponding extremal entropy (i.e., by the entropy pertaining to the
uniquely defined extremal limit $\lim_{T\rightarrow 0^{+}}$ of the $T$%
-dependent near-extremal entropy); actually, this breakdown is expected, as
the near-extremal entropy is no more a homogeneous function of degree two of
the charges \cite{Marrani:2017ihg,Marrani:2019zsn}.

In the present paper, we will show how to circumvent such an obstruction,
and thus how to consistently formulate a Freudenthal duality for the entropy
of near-extremal black holes; we anticipate that the price to pay for it is
a (remarkably, non-linear and uniquely defined) \textit{transformation of
the temperature} $T$ itself.\bigskip

The plan of the paper is as follows.

In Sec. \ref{sec:review} we review some topics of the framework in which we
will derive our results, such as $\mathcal{N}=2$, $D=4$ ungauged
supergravity (or, more in general, Maxwell-Einstein theories coupled to a
non-linear sigma model of scalar fields in absence of a potential) in Sec. %
\ref{Basics} and the approach to the entropy of near-extremal black holes
based on Jackiw-Teitelboim gravity in Sec. \ref{JT}. We will also comment on
the invariance of extremal entropy under Freudenthal duality and on the
non-invariance of near-extremal entropy under a na\"{\i}ve notion of
Freudenthal duality in presence of a non-vanishing temperature, respectively
in Secs. \ref{T=0-inv} and \ref{T-neq-0}. Then, in Sec. \ref{NEFD} we
present near-extremal Freudenthal duality, which is the consistent
generalization of the intrinsically non-linear Freudenthal duality map
(originally introduced for extremal black holes) to the class of
near-extremal black hole solutions of Einstein-Maxwell equations of motion.
In particular, the introduction of a non-linear transformation of the
temperature is discussed in Sec. \ref{deltaT}, whereas its uniqueness and
identification with a specific solution of a quartic algebraic equation are
discussed in Secs. \ref{Uniqueness} and \ref{Precision}, respectively.
In Sec. \ref{stu} we briefly present an application of our results to the STU model.
Concluding remarks and hints for further developments are given in the final
Sec. \ref{Conclusion}. Two appendices conclude the paper, dedicated to the
statement of the technical results which we will exploit in order to derive
our results: namely, \textit{Descartes' rule of signs} (in App. \ref%
{app:descarte}) and \textit{Sturm's Theorem} (in App. \ref{app:sturm}).

\section{Freudenthal duality in $\mathcal{N}=2$, $D=4$ ungauged supergravity}

\label{sec:review}

\subsection{A few basics}

\label{Basics}

Here we will briefly review the observation made by two of the present
authors in \cite{Chattopadhyay:2021vog}, which dealt with the Freudenthal
duality of a near-extremal black hole in type $IIA$, $\mathcal{N}=2$, $D=4$
ungauged supergravity, by exploiting the JT gravity theory. The bosonic part
of the corresponding \textit{ungauged} supergravity action coupled with an
arbitrary number of vector multiplets can be written as
\begin{eqnarray}  \label{eq:iiaction}
\mathcal{S}&=&-\frac{1}{8\pi G_{4}}\int d^{4}x\sqrt{-G}\Bigg( -{\frac{R}{2}}%
+h_{a\bar{b}}\partial_{\mu }x^{a}\partial _{\nu }\bar{x}^{\bar{b}}G^{\mu\nu }
\notag \\
&&-\mu _{\Lambda \Sigma }\mathcal{F}_{\mu \nu }^{\Lambda }\mathcal{F}%
_{\lambda \rho }^{\Sigma }G^{\mu \lambda }G^{\nu \rho }-\nu _{\Lambda \Sigma}%
\mathcal{F}_{\mu \nu }^{\Lambda }\ast \mathcal{F}_{\lambda \rho
}^{\Sigma}G^{\mu \lambda }G^{\nu \rho }\Bigg),
\end{eqnarray}%
with $G_{4}$ standing for the four-dimensional Newton's constant, and $%
G_{\mu \nu }$, $R$ and $G$ respectively denoting the space-time metric, the
Ricci scalar and the determinant of the metric. On the other hand, $h_{a\bar{%
b}}$ denotes the moduli space metric for the $n$ complex scalars $x^{a}$.
The action (\ref{eq:iiaction}) also includes $n+1$ 1-form Maxwell potentials
$A_{\mu }^{\Lambda }$, $\Lambda =0,1,...,n$, whose corresponding 2-form
field strengths are denoted as $\mathcal{F}^{\Lambda }$.

The action (\ref{eq:iiaction}) can be conceived as the purely bosonic sector
of a Maxwell-Einstein supergravity theory arising as the large-volume,
low-energy limit of type $IIA$ string theory compactified on some Calabi-Yau
threefold $\mathcal{M}$. For this theory, the prepotential $F$ and the K\"{a}%
hler potential $K$ read as
\begin{equation}
F=D_{abc}{\frac{X^{a}X^{b}X^{c}}{X^{0}}},\quad K=-\log \left[ i\sum_{\Lambda
=0}^{n}\left( \overline{X^{\Lambda }}\partial _{\Lambda }F-X^{\Lambda }%
\overline{\partial _{\Lambda }F}\right) \right] ,  \label{eq:prepot_kahler}
\end{equation}%
where $X^{a}=x^{a}X^{0}$. The K\"{a}hler potential $K$ yields the moduli
space metric $h_{a\bar{b}}=\partial _{a}\partial _{\bar{b}}K$, where the
partial derivatives are taken with respect to the complex moduli fields. The
gauge coupling constants in (\ref{eq:iiaction}) can be considered to be
given by the real and imaginary part of a complex symmetric matrix $\mathcal{%
N}_{\Lambda \Sigma }$ as $\mu =\text{Im}\mathcal{N}$ and $\nu =-\text{Re}%
\mathcal{N}$ with
\begin{equation}
\mathcal{N}_{\Lambda \Sigma }=\overline{F_{\Lambda \Sigma }}+2i{\frac{\text{%
Im}F_{\Lambda \Omega }\text{Im}F_{\Sigma \Pi }X^{\Omega }X^{\Pi }}{\text{Im}%
F_{\Omega \Pi }X^{\Omega }X^{\Pi }}};\quad F_{\Lambda \Sigma }:=\partial
_{\Lambda }\partial _{\Sigma }\,F.  \label{formN}
\end{equation}%
Defining the 2-forms $\alpha _{a}$ to be the basis of the second integral
cohomology group $H^{2}(\mathcal{M},\mathbb{Z})$, the triple intersection
numbers $D_{abc}$ of $\mathcal{M}$ are defined as
\begin{equation}
D_{abc}:=\frac{1}{6}\int_{\mathcal{M}}\alpha _{a}\wedge \alpha _{b}\wedge
\alpha _{c}.
\end{equation}%
\bigskip

\textbf{Nota Bene :} while we refer to type $IIA$, $\mathcal{N}=2$, $D=4$
ungauged supergravity for a concrete framework, all results obtained in this
paper actually hold irrespective of (local) supersymmetry, namely they
actually, hold for any Maxwell-Einstein theory in $D=4$ space-time
dimensions with an action given by (\ref{eq:iiaction}). This is a remarkable
fact, which was proved within the general treatment firstly given in \cite%
{Ferrara:2011gv}, and then developed and detailed in subsequent works, such
as \cite{Borsten:2012pd}, \cite{Marrani:2015wra}, \cite{Marrani:2017ihg} and
\cite{Marrani:2019zsn}.

\subsection{\label{JT}Jackiw-Teitelboim gravity and near-extremal entropy}

Before discussing the application of Freudenthal duality to near-extremal
black hole, here we will briefly summarize the procedure to derive the
near-extremal black hole entropy; further details and comments can be found
in \cite{Iliesiu:2020qvm,Chattopadhyay:2021vog}. A crucial property is the
fact that the near-horizon geometry of an asymptotically flat, static,
spherically symmetric, dyonic extremal black hole factorises as $%
AdS_{2}\otimes S^{2}$, with both factors having the same radius; this
results in the so-called Bertotti-Robinson geometry \cite%
{Bertotti:1959pf,Robinson:1959ev}. We will consider the simple case, in
which the near-extremal black hole inherits the spherical symmetry and the
staticity from its extremal counterpart/limit. In this scenario, one can
split the computation of the near-extremal black hole entropy into four
steps, briefly listed below \cite{Iliesiu:2020qvm,Chattopadhyay:2021vog}.

\begin{enumerate}
\item[I] \textbf{Dimensional reduction: }The first step is to start with a
dimensional reduction, by considering a spherically symmetric \textit{Ansatz}%
\footnote{%
We refer e.g. to \cite{Iliesiu:2020qvm} for a more general \textit{Ansatz}
allowing for a stationary rotation, with non-zero angular momentum.} for the
four-dimensional, asymptotically flat black hole metric $G_{\mu \nu }$ as
\begin{equation}
ds^{2}=G_{\mu \nu }dx^{\mu }dx^{\nu }=\tilde{g}_{\alpha \beta }dx^{\alpha
}dx^{\beta }+{\Phi }^{2}d\Omega _{2}^{2};\quad \mu ,\nu \in \lbrack
0,3],\quad \alpha ,\beta \in \lbrack 0,1].  \label{dimredanst}
\end{equation}%
Furthermore, we denote the electric and magnetic charges of the black hole
by $Q_{\Lambda }$ and $P^{\Lambda }$ respectively, and by spherical symmetry
we find that $\mathcal{F}_{\theta \phi }^{\Lambda }=P^{\Lambda }sin\theta $.
Within this set of assumptions, a straightforward dimensional reduction
would generally contain derivatives of $\Phi $, which can be regarded as an
additional scalar field (\textit{dilaton}). As we are aiming at getting a
JT-like action, the derivatives of $\Phi $ need to be removed. In order to
achieve this, one can perform a Weyl-rescaling of the metric, as follows
\begin{equation}
g_{\alpha \beta }={\frac{\Phi }{\Phi _{0}}}\tilde{g}_{\alpha \beta }.
\label{Weyl-r}
\end{equation}%
Therefore, the dimensionally reduced and Weyl-rescaled version of (\ref%
{eq:iiaction}) turns out to be \cite{Chattopadhyay:2021vog}
\begin{align}  \label{eq:fulldimred}
\mathcal{\hspace*{0pt}S}_{tot}=& {\frac{1}{2G_{4}}}\int d^{2}x\sqrt{-g}\,%
\Bigg[ {\frac{\Phi ^{2}R(g)}{2}}+{\frac{\Phi _{0}}{\Phi }}+{\frac{\Phi ^{3}}{%
\Phi _{0}}}\mu _{\Lambda \Sigma }\mathcal{F}_{\alpha \beta }^{\Lambda }%
\mathcal{F}^{\Sigma \,\alpha \beta }+\frac{2\Phi _{0}}{\Phi ^{3}}\mu
_{\Lambda \Sigma }P^{\Lambda }P^{\Sigma }  \notag \\
&-\Phi ^{2}\,h_{a\bar{b}}\partial _{\alpha }x^{a}\partial _{\beta }\bar{x}^{%
\bar{b}}g^{\alpha \beta }\Bigg] +{\frac{1}{2G_{4}}}\int d^{2}x(2\nu
_{\Lambda \Sigma }P^{\Lambda }\mathcal{F}_{\alpha \beta }^{\Sigma }\epsilon
^{\alpha \beta })  \notag \\
&+16\pi \int dt\,\sqrt{-h}\left[ {\frac{\Phi ^{3}}{\Phi _{0}}}\mu _{\Lambda
\Sigma }\,n_{\alpha }\mathcal{F}^{\Lambda \,\alpha \beta }A_{\beta }^{\Sigma
}+{\frac{1}{\sqrt{-g}}}\,\nu _{\Lambda \Sigma }\,n_{r}P^{\Lambda
}A_{t}^{\Sigma }\right]  \notag \\
& +{\frac{1}{2G_{4}}}\pi \int dt\sqrt{-h}\,\Phi ^{2}\,k,
\end{align}%
where $g$ is the determinant of the two-dimensional metric (\ref{Weyl-r}),
and in the last term, we have also included the dimensionally reduced
Gibbons-Hawking-York (GHY) boundary term, with $k$ being the extrinsic
curvature of the induced boundary metric after dimensional reduction. The
two dimensional Levi-Civita symbol $\epsilon ^{\alpha \beta }$ is taken such
that $\epsilon ^{rt}=1$.

\item[II] \textbf{Deriving the effective dilaton action:} Next, we will
further introduce some simplifying assumptions, namely focusing on
double-extremal solutions, in which the moduli fields are constant in
space-time. Consequently, the equations of motion for moduli fields get
trivially satisfied, and one is only left with the equations of motion for
the dilaton $\Phi $, the metric $g_{\alpha \beta }$ (\ref{Weyl-r}) and the
gauge fields $A^{\Lambda }$. To get a JT-like gravity action, one then has
to integrate out the gauge fields from the equations by inserting the gauge
field solutions (along with the proper scaling, due to the Weyl
transformation (\ref{Weyl-r}), as discussed in \cite{Chattopadhyay:2021vog}.
By doing so, the equations of motion for $\Phi $ and $g_{\alpha \beta }$
boil down to
\begin{equation}
\begin{split}
\Phi R(g)-{\frac{\Phi _{0}}{\Phi ^{2}}}-6{\frac{\Phi _{0}}{\Phi ^{4}}}\left(
\mathcal{X}+\frac{G_{4}}{2}\mu _{\Lambda \Sigma }p^{\Lambda }p^{\Sigma
}\right) & =0; \\
1+{\frac{2}{\Phi ^{2}}}\left( \mathcal{X}+\frac{G_{4}}{2}\mu _{\Lambda
\Sigma }p^{\Lambda }p^{\Sigma }\right) +{\frac{\Phi }{\Phi _{0}}}(\nabla
_{\alpha }\Phi )^{2}+{\frac{\Phi ^{2}}{\Phi _{0}}}\nabla ^{2}\Phi & =0,
\end{split}
\label{eq:2deom}
\end{equation}%
where
\begin{eqnarray}
\mathcal{X}(\mu, \nu, P, Q):&=&\frac{G_{4}}{2}\Bigg( (\mu^{-1})^{\Lambda
\Sigma }q_{\Lambda }q_{\Sigma }+p^{\Sigma }{(\nu \mu^{-1})_{\Sigma }}%
^{\Lambda }q_{\Lambda }+q_{\Lambda }{(\mu ^{-1}\nu)^{\Lambda }}_{\Sigma
}p^{\Sigma }  \notag \\
&&+p^{\Lambda }(\nu \mu ^{-1}\nu )_{\Lambda\Sigma }p^{\Sigma }\Bigg).
\end{eqnarray}
Crucially, the same equations of motion (\ref{eq:2deom}) can be obtained as
Euler-Lagrange equations from the effective, two-dimensional action
\begin{eqnarray}  \label{eq:2deffac}
\mathfrak{S}&=&-{\frac{1}{2G_{4}}}\int d^{2}x\,\sqrt{-g}\left[-{\frac{%
\Phi^{2}R}{2}}-{\frac{\Phi _{0}}{\Phi }}-{\frac{2\Phi_{0}}{\Phi ^{3}}}\left(%
\mathcal{X}+\frac{G_{4}}{2}\mu_{\Lambda \Sigma }p^{\Lambda
}p^{\Sigma}\right) \right]  \notag \\
&&+{\frac{1}{2G_{4}}}\int dt\sqrt{-h}\Phi ^{2}k,
\end{eqnarray}%
which can simply be conceived as a two-dimensional dilatonic gravity theory,
\begin{equation}
\mathfrak{S}=-{\frac{1}{2G_{4}}}\int d^{2}x\,\sqrt{-g}\left[ -{\frac{\Phi
^{2}R}{2}}-U(\Phi )\right] +{\frac{1}{2G_{4}}}\int dt\sqrt{-h}\Phi ^{2}K,
\label{eq:altenrate2d}
\end{equation}%
with the \textquotedblleft effective dilaton potential\textquotedblright\
defined as%
\begin{equation}
{U(\Phi ):={\frac{\Phi _{0}}{\Phi }}+{\frac{2\Phi _{0}}{\Phi ^{3}}}\left(
\mathcal{X}+\frac{G_{4}}{2}\mu _{\Lambda \Sigma }p^{\Lambda }p^{\Sigma
}\right) .\label{dil-pot}}
\end{equation}

\item[III] \textbf{Extremal black hole entropy from the effective action:}
Following \cite{Iliesiu:2020qvm,Chattopadhyay:2021vog}, the most general
static solution for the dilatonic gravity theory (\ref{eq:altenrate2d})-(\ref%
{dil-pot}) can be written as%
\begin{equation}
\Phi =r,  \label{metr-1}
\end{equation}%
thus yielding that%
\begin{eqnarray}
ds^{2} &=&\frac{\Phi }{\Phi _{0}}\left( -f(r)dt^{2}+\frac{dr^{2}}{f(r)}%
\right) , \\
f(r) &:&=1+\frac{\mathcal{C}}{\Phi }+\frac{2\Theta (\mu ,\nu ,P,Q)}{\Phi ^{2}%
}, \\
\Theta (\mu ,\nu ,P,Q) &=&:-\left( \mathcal{X}(\mu ,\nu ,P,Q)+\frac{G_{4}}{2}%
\mu _{\Lambda \Sigma }p^{\Lambda }p^{\Sigma }\right) ,  \label{metr-2}
\end{eqnarray}%
where $\Theta $ is nothing but (one half of) the \textquotedblleft black
hole effective potential\textquotedblright\ \cite{Ferrara:1997tw}.
Furthermore, $\mathcal{C}$ is an integration constant, which needs to be
fixed from the boundary conditions. As mentioned above, the whole space-time
metric $G_{\mu \nu }$ is well approximated by\ the $AdS_{2}\otimes S^{2}$
Bertotti-Robinson geometry in the near-horizon (NH) region, in which $%
r-r_{0}\ll r_{0}$, with $r_{0}$ denoting the radius of the (unique) event
horizon of the extremal black hole. Both for extremal and near-extremal
black holes, the far-horizon (FH) region still remains well approximated by
the extremal metric of the form
\begin{equation}
ds_{FH}^{2}=-f_{FH}(r)\,dt^{2}+{\frac{dr^{2}}{f_{FH}(r)}}+r^{2}\,d\Omega
_{2}^{2}.
\end{equation}%
On the other hand, as from the above discussion, in the NH region the metric
of the extremal black hole reads as
\begin{eqnarray}
\quad ds^{2} &=&-\frac{(r-r_{0})^{2}}{L_{2}^{2}}dt^{2}+\frac{L_{2}^{2}}{%
(r-r_{0})^{2}}dr^{2}+r_{0}^{2}d\Omega _{2}^{2},  \notag \\
\text{with~}L_{2} &=&r_{0}.  \label{metr-extr}
\end{eqnarray}%
One can similarly take the NH limit of the metric (\ref{metr-1})-(\ref%
{metr-2}), whose comparison with (\ref{metr-extr}) allows one to fix
\begin{equation}
\mathcal{C}=-2r_{0},\quad \text{and }\quad \Phi =r_{0}.
\end{equation}%
Since we are considering an asymptotically flat, four-dimensional
space-time, without any loss of generality we can set $L_{2}=\Phi _{0}=r_{0}$%
, with $L_{2}$ being the $AdS_{2}$ radius. At the (unique) event horizon $%
r=r_{0}$ of the extremal black hole, the equations of motion (\ref{eq:2deom}%
) imply that
\begin{equation}
r_{0}^{2}=2\Theta ^{\ast }\left( P,Q\right) ,\quad R=-{\frac{2}{\Phi _{0}^{2}%
}},  \label{eq:dilatonatheorizon}
\end{equation}%
where $\Theta ^{\ast }\left( P,Q\right) :=\Theta (\mu _{H}\left( P;Q\right)
,\nu _{H}\left( P,Q\right) ,P,Q)$, where $\mu _{H}$ and $\nu _{H}$ are the
purely $P,Q$-dependent matrices $\mu $ and $\nu $ at\footnote{%
As it is well known, at $r=r_{0}$ the \textit{attractor mechanism} \cite%
{Ferrara:1995ih,Ferrara:1996dd,Ferrara:1996um,Ferrara:1997tw} takes place,
fixing the horizon values of the complex moduli in terms of the e.m. charges
only : $x_{H}^{a}=x_{H}^{a}(P,Q)$, assuming no flat directions (for
instance, we have in mind BPS extremal black holes in $\mathcal{N}=2$
supergravity). Thus, the matrices $\mu =\mu (x,\bar{x})$ and $\nu =\nu (x,%
\bar{x})$ at $r=r_{0}$ become purely dependent on the e.m. charges : $\mu
_{H}=\mu (x_{H}\left( P,Q\right) ,\bar{x}_{H}\left( P,Q\right) )\equiv \mu
_{H}\left( P,Q\right) $ and $\nu _{H}=\nu (x_{H}\left( P,Q\right) ,\bar{x}%
_{H}\left( P,Q\right) )\equiv \nu _{H}\left( P,Q\right) $. Rigorously
speaking, it should be noted that for an ungauged Einstein-Maxwell-scalar
theory with non-trivial e.m. duality and \textit{non-homogeneous} scalar
manifold, the holding of the Attractor Mechanism at the horizon of the class
of extremal black holes under consideration should indeed be assumed, and
not necessarily understood. As mentioned above, in this manuscript, we had in mind
extremal black hole solutions exhibiting attractor mechanism and displaying
no flat directions, such as $\mathcal{N}=2$ ($1/2$-)BPS extremal black holes.}
 $r=r_{0}$. In the NH region, the actual values of the dilaton and the
metric can be thought of as small perturbations around the background
solution. Interestingly, the \textit{extremal} black hole entropy can be
recovered from these background solutions, whereas perturbations to the
above scenario give rise to \textit{near-extremal} corrections. In the
conformal gauge, the background solution can be written as
\begin{equation}
ds^{2}=e^{2\omega _{0}}(-dt^{2}+d\rho ^{2}),\quad e^{2\omega _{0}}={\frac{%
L_{2}^{2}}{\rho ^{2}}},
\end{equation}%
from which one can calculate that
\begin{equation}
R=-2e^{-2\omega }\left( {\frac{\partial ^{2}\omega }{\partial \rho ^{2}}}-{%
\frac{\partial ^{2}\omega }{\partial t^{2}}}\right) ,\quad \text{and}\quad k=%
{\frac{1}{r_{0}}},
\end{equation}%
obtained after switching to the Euclidean time $\tau :=it$. In fact, the
Euclidean setup simplifies the calculation very much, eventually yielding
the extremal entropy to be given by
\begin{equation}
S_{0}=\frac{\pi r_{0}^{2}}{G_{4}}={\frac{2\pi \lvert \Theta ^{\ast }\left(
P,Q\right) \rvert }{G_{4}}}.  \label{eq:s0}
\end{equation}%
Following \cite{Chattopadhyay:2021vog}, one can show that the result (\ref%
{eq:s0}) is the same as the one obtained some decades ago by Shmakova after
incorporating the attractor values of the double-extremal $STU$ model \cite%
{Shmakova:1996nz}.

\item[IV] \textbf{The near-extremal black hole entropy:} Remarkably, in the
scenario under consideration, a \textit{near-extremal} black hole can be
obtained simply by \textquotedblleft perturbating\textquotedblright\ an
\textit{extremal} black hole \cite{Iliesiu:2020qvm,Chattopadhyay:2021vog}.
Therefore, to calculate the near-extremal black hole entropy we will now
switch some perturbations on the dilaton and the metric background
solutions, denoted by $\phi$ and $\Omega$, respectively,
\begin{equation}
\Phi =r_{0}(1+\phi ),~~~~~~~\omega =\omega _{0}+\Omega .  \label{eq:fluc}
\end{equation}%
As retrieved at point III above, the leading order part of the action, along
with the leading order GHY term, gives rise to the extremal black hole
entropy. On the other hand, the first-order perturbation of the action
(without the boundary term) vanishes, due to the equations of motion. One is
thus left with the first-order term in perturbation only coming out from the
GHY term, as
\begin{equation}
\mathcal{S}_{1,~\text{boundary}}=-\frac{r_{0}^{2}}{G_{4}}\int dx\,\sqrt{h}%
\phi K,  \label{eq:ghyaction}
\end{equation}%
where the subscript \textquotedblleft $1$\textquotedblright\ denotes the
evaluation at the first order on perturbation theory. Thus, the evaluation
of the integral (\ref{eq:ghyaction}) will yield the correction to the
near-extremal black hole entropy (with respect to the entropy of the
corresponding extremal black hole limit), namely the correction at the first
order in perturbation theory to the extremal black hole itself. Following
\cite{Chattopadhyay:2021vog}, one can evaluate\footnote{%
We here disregard all the intricacies and technicalities in the evaluation
of the integral (\ref{eq:ghyaction}), since that is not the purpose of this
paper.} the integral and obtain the near-extremal correction to the extremal
black hole entropy as
\begin{equation}
\delta S=\frac{4\pi ^{2}}{G_{4}}r_{0}^{3}T.  \label{exen}
\end{equation}
\end{enumerate}

\subsection{\label{T=0-inv}$T=0$ : invariance of \textit{extremal} entropy}

Freudenthal duality is a symmetry of the entropy $S_{0}$ of asymptotically
flat, static, spherically symmetric, dyonic extremal black holes in (not
necessarily supersymmetric) Maxwell-Einstein-scalar theories in four
space-time dimensions which have a non-trivial e.m. duality symmetry,
endowed with a symplectic structure satisfying the identity defining
\textquotedblleft generalized special geometry\textquotedblright , given
e.g. by Eqs. (1.14) and (1.15) of \cite{Ferrara:2011gv}. It is here worth
pointing out that in the treatment of \cite{Ferrara:2011gv} no constraints
on the special K\"{a}hler geometry of $\mathcal{N}=2$ (vector multiplets')
scalar manifolds are considered. However, as
discussed in footnote 2, we have assumed that, up to some flat directions
(whose interplay with Freudenthal duality is discussed in\ \cite%
{Ferrara:2013zga}, and - in a slightly less general framework - in \cite%
{Fernandez-Melgarejo:2013ksa}), the scalar fields only have attractor
directions at the event horizon of the extremal black hole (namely, the
Attractor Mechanism holds, up to some flat directions).

By defining the Freudenthal transformation acting on the e.m. black hole
charges as%
\begin{equation}
\mathcal{\tilde{Q}}^{M}:=\Omega ^{MN}{\frac{\partial S_{0}(\mathcal{Q})}{%
\partial \mathcal{Q}^{N}}=}\mathcal{\tilde{Q}}^{M}\left( \mathcal{Q}\right) ,
\label{F-S0}
\end{equation}%
it holds that \cite{Borsten:2009zy, Ferrara:2011gv}%
\begin{equation}
S_{0}(\mathcal{\tilde{Q}})=S_{0}(\mathcal{Q}).  \label{inv-S0}
\end{equation}%
In this framework, the Freudenthal map (\ref{F-S0}) is \textit{%
anti-involutive} since it can be recast in the following form:%
\begin{equation}
\mathcal{\tilde{Q}}^{M}=-\Omega ^{MN}\mathcal{M}_{H|NP}\left( \mathcal{Q}%
\right) \mathcal{Q}^{P},  \label{F-S0-2}
\end{equation}%
where $\mathcal{M}_{H}(\mathcal{Q})$ denotes the matrix $\mathcal{M}$ of
special K\"{a}hler geometry (see e.g. \cite{Ferrara:2013zga} and Refs.
therein), evaluated at the (unique) event horizon of the extremal black hole
(at which the \textit{attractor mechanism} \cite%
{Ferrara:1995ih,Ferrara:1996dd,Ferrara:1996um,Ferrara:1997tw} takes place;
see also ). The $\mathcal{M}_{H}(\mathcal{Q})$ is real, symmetric and
symplectic:%
\begin{eqnarray}
\mathcal{M}_{H}^{T} &=&\mathcal{M}_{H}; \\
\overline{\mathcal{M}_{H}} &=&\mathcal{M}_{H}; \\
\mathcal{M}_{H}\Omega \mathcal{M}_{H} &=&\Omega .
\end{eqnarray}%
These properties, together with $\Omega ^{2}=-\mathbb{I}$, imply the
anti-involutivity of the Freudenthal map defined by (\ref{F-S0}), or,
equivalently, by (\ref{F-S0-2}) :%
\begin{equation}
\widetilde{\mathcal{\tilde{Q}}}^{M}:=-\mathcal{Q}^{M}.
\end{equation}

In the following treatment, we will present a consistent generalization of
the definition (\ref{F-S0}) of the Freudenthal duality map, in presence of a
\textit{non-vanishing temperature} $T\neq 0$, namely for \textit{%
near-extremal} black holes. It is, however, worth noticing that such a
variant of Freudenthal duality, named \textit{near-extremal Freudenthal
duality}, is no more anti-involutive.

\subsection{\label{T-neq-0}$T\neq 0$ : no invariance of \textit{near-extremal%
} entropy}

Thus, within the framework under consideration, the near-extremal black hole
entropy (denoted by the subscript \textquotedblleft $NE$\textquotedblright )
reads\footnote{%
We will henceforth understand the black hole entropy in units of $\pi $.}
\begin{equation}
S_{NE}\left( \mathcal{Q};T\right) \simeq S_{0}\left( \mathcal{Q}\right)
+\delta S\left( \mathcal{Q};T\right) =\frac{\pi r_{0}^{2}}{G_{4}}\left(
1+4\pi Tr_{0}\right) =S_{0}+4\sqrt{\pi G_{4}}\,\,TS_{0}^{3/2}.
\label{eq:fullS}
\end{equation}%
It should be remarked that in formula (\ref{eq:fullS}) we have disregarded
the logarithmic term arising from the \textit{quantum corrections} due to
the measure of the partition function \cite%
{Iliesiu:2020qvm,Chattopadhyay:2021vog}; in other words, we are focusing
only on the semi-classical part of the near-extremal entropy; see also the
concluding remarks in Sec. \ref{Conclusion}.

Starting from (\ref{eq:fullS}), one can then easily check that $S_{NE}$ is
not invariant under the na\"{\i}ve definition of Freudenthal duality,
\begin{equation}
\hat{\mathcal{Q}}^{M}:=\Omega ^{MN}{\frac{\partial S_{NE}(\mathcal{Q};T)}{%
\partial \mathcal{Q}^{N}}=}\hat{\mathcal{Q}}^{M}\left( \mathcal{Q};T\right) ,
\label{Q-hat}
\end{equation}%
where $\Omega $ is the invariant metric of the symplectic representation
space of e.m. black hole charges spanned by $\mathcal{Q}$. Namely, one can
check that \cite{Chattopadhyay:2021vog}
\begin{equation}
S_{NE}(\hat{\mathcal{Q}};T)\neq S_{NE}(\mathcal{Q};T).  \label{non-inv}
\end{equation}%
Moreover, it can be further shown that the invariance of the near-extremal
black hole entropy $S_{NE}$ is not restored, also when including the
aforementioned quantum corrections (by means of a logarithmic term) in $%
S_{NE}$ itself \cite{Chattopadhyay:2021vog}.

\section{\label{NEFD}$T\neq 0$ : N\textit{ear-extremal} Freudenthal duality}

In this section, we will generalise the notion of Freudenthal duality for a
\textit{near-extremal} black hole, namely in presence of a \textit{%
non-vanishing} temperature $T\neq 0$. From the discussion in Sec. \ref%
{T-neq-0}, Eq. (\ref{non-inv}) means that two near-extremal black holes
(having the same temperature) cannot have the same entropy if their e.m.
charges are related by the Freudenthal duality map constructed from the
full-fledged near-extremal entropy (\ref{eq:fullS}); trivially, as observed
in \cite{Chattopadhyay:2021vog}, the \textit{near-extremal} entropy is
invariant only under the Freudenthal map constructed from the \textit{%
extremal} entropy.

Still, it might be possible that two near-extremal black holes are
Freudenthal dual to each other (with the Freudenthal map defined by the
near-extremal entropy (\ref{eq:fullS})), that they have the same entropy,
but then this should imply that the two near-extremal black holes have
\textit{different temperatures}. The present section will be devoted to
considering this possibility in detail. So, we will be aiming at varying the
temperature such that two Freudenthal dual near-extremal black holes have
the same entropy.

For simplicity's sake, we report here Eq. (\ref{eq:fullS}) as

\begin{equation}
S_{NE}(\mathcal{Q};T)=S_{0}\left( \mathcal{Q}\right) +\alpha
TS_{0}^{3/2}\left( \mathcal{Q}\right) ,  \label{eq:newnonen}
\end{equation}%
where $\alpha :=4\sqrt{\pi G_{4}}$, such that $S_{NE}(\mathcal{Q}%
;T)\geqslant 0~\forall \mathcal{Q},T$, where\footnote{%
We assume that the extremal limit is well-defined and unique.}
\begin{equation}
S_{0}\left( \mathcal{Q}\right) :=\lim_{T\rightarrow 0^{+}}S_{NE}(\mathcal{Q}%
;T),  \label{eq:s0def}
\end{equation}%
is the corresponding extremal black hole entropy.

Let us now define the \textit{near-extremal (on-shell) Freudenthal duality}
(acting on black hole e.m. charges $\mathcal{Q}$), namely, let us make the
definition (\ref{Q-hat}) explicit by specifying (\ref{eq:newnonen})
\begin{eqnarray}  \label{F_T}
\mathfrak{F}_{T}\left( \mathcal{Q}\right) \equiv \hat{\mathcal{Q}}\left(
\mathcal{Q};T\right) :&= &\Omega \frac{\partial S_{NE}(\mathcal{Q};T)}{%
\partial \mathcal{Q}}=\Omega \frac{\partial S_{0}\left( \mathcal{Q}\right) }{%
\partial \mathcal{Q}}+\frac{3\alpha }{2}T\sqrt{S_{0}\left( \mathcal{Q}%
\right) }\Omega \frac{\partial S_{0}\left( \mathcal{Q}\right) }{\partial
\mathcal{Q}}  \notag \\
&=&\left( 1+\frac{3\alpha }{2}T\sqrt{S_{0}\left( \mathcal{Q}\right) }\right)
\Omega \frac{\partial S_{0}\left( \mathcal{Q}\right) }{\partial \mathcal{Q}}
\notag \\
&=&\left( 1+\frac{3\alpha }{2}T\sqrt{S_{0}\left( \mathcal{Q}\right) }\right)
\mathfrak{F}_{0}\left( \mathcal{Q}\right) ,
\end{eqnarray}%
where we have recalled the definition (\ref{F-S0}),
\begin{equation}
\mathfrak{F}_{0}\left( \mathcal{Q}\right) \equiv \mathcal{\tilde{Q}}\left(
\mathcal{Q}\right) :=\Omega \frac{\partial S_{0}\left( \mathcal{Q}\right) }{%
\partial \mathcal{Q}}=\lim_{T\rightarrow 0^{+}}\mathfrak{F}_{T}\left(
\mathcal{Q}\right) \equiv \lim_{T\rightarrow 0^{+}}\hat{\mathcal{Q}}\left(
\mathcal{Q};T\right) .  \label{F0}
\end{equation}%
The action of the near-extremal Freudenthal duality map (\ref{F_T}) on the
near-extremal black hole entropy (\ref{eq:newnonen}) can then be written as
\begin{eqnarray}
\mathfrak{F}_{T}\left( S_{NE}(\mathcal{Q};T)\right) &=&S_{NE}(\mathfrak{F}%
_{T}\left( \mathcal{Q}\right) ;T)=S_{0}\left( \mathfrak{F}_{T}\left(
\mathcal{Q}\right) \right) +\alpha TS_{0}^{3/2}\left( \mathfrak{F}_{T}\left(
\mathcal{Q}\right) \right)  \notag \\
&=&\left( 1+\frac{3\alpha }{2}T\sqrt{S_{0}\left( \mathcal{Q}\right) }\right)
^{2}\left[ 1+\alpha T\left( 1+\frac{3\alpha }{2}T\sqrt{S_{0}\left( \mathcal{Q%
}\right) }\right) \sqrt{S_{0}\left( \mathcal{Q}\right) }\right] S_{0}\left(
\mathcal{Q}\right) ,  \notag \\
&&  \label{transf-S}
\end{eqnarray}%
where we have used the homogeneity of degree two of the extremal entropy $%
S_{0}\left( \mathcal{Q}\right) $ as a function of $\mathcal{Q}$,%
\begin{equation}
S_{0}\left( \lambda \mathcal{Q}\right) =\lambda ^{2}S_{0}\left( \mathcal{Q}%
\right) ,~\forall \lambda \in \mathbb{R}_{0},
\end{equation}%
as well as its invariance under $\mathfrak{F}_{0}\left( \mathcal{Q}\right) $
(\ref{F0}), expressed by Eq. (\ref{inv-S0}) \cite{Ferrara:2011gv}.

In Sec. \ref{T-neq-0} we have recalled the result (\ref{non-inv}) obtained
in \cite{Chattopadhyay:2021vog}, two near-extremal black holes, with charges
related by the Freudenthal map (\ref{Q-hat}) (or, more explicitly, (\ref{F_T}%
)) and with same temperature $T\neq 0$, \textit{cannot} have the same
entropy. By virtue of Eq. (\ref{transf-S}), this fact can be retrieved by
observing that the condition for invariance of $S_{NE}(\mathcal{Q};T)$ under
the Freudenthal map (\ref{F_T}), namely
\begin{equation}
\mathfrak{F}_{T}\left( S_{NE}(\mathcal{Q};T)\right) =S_{NE}(\mathcal{Q};T),
\label{cond-inv}
\end{equation}%
can be recast into the algebraic inhomogeneous equation of degree three

\begin{equation}
x^{3}+2x^{2}+2x+\frac{8}{9}=0  \label{eq:cubiceq}
\end{equation}%
by exploiting Eqs. (\ref{eq:newnonen}) and (\ref{transf-S}), and defining%
\begin{equation}
x:=\alpha T\sqrt{S_{0}(Q)}.  \label{x}
\end{equation}%
By recalling that $\alpha :=4\sqrt{\pi G_{4}}$, the condition of invariance (%
\ref{cond-inv}) has a physically meaningful solution (which determines the
analytical functional form of $T=T\left( S_{0}(\mathcal{Q})\right) $)
\textit{iff} there exists a positive solution to Eq. (\ref{eq:cubiceq}).
However, as obtained in \cite{Chattopadhyay:2021vog}, there are no positive
solutions to Eq. (\ref{eq:cubiceq}). Indeed, by exploiting \textit{%
Descartes' rule of signs} (discussed in App. \ref{app:descarte}), one can
show the cubic inhomogeneous Eq. (\ref{eq:cubiceq}) \textit{does not admit
any positive real root}. Equivalently, the direct resolution of Eq. (\ref%
{eq:cubiceq}) yields one real, negative root and two complex conjugate roots%
\footnote{%
The exact numerical values of the roots are $x=\{-0.87,\,-0.56+0.84i,%
\,-0.56-0.84i\}$.}. Given the physical meaning of $x$ defined by (\ref{x}),
these corresponding solutions are unphysical, thereby reinforcing the
results of \cite{Chattopadhyay:2021vog}, as anticipated.

\subsection{\label{deltaT}$T\neq 0$ and $\protect\delta T\neq 0$ :
invariance of \textit{near-extremal} entropy}

As anticipated above, not having any positive real roots for the cubic
inhomogeneous Eq. (\ref{eq:cubiceq}) prompts us to consider a transformation
of the temperature $T$, which we assume to be given by
\begin{equation}
T\longrightarrow T+\delta T,  \label{transf-T}
\end{equation}%
with%
\begin{equation}
\delta T=\delta T\left( S_{0}(\mathcal{Q}),T\right) .  \label{delta-T}
\end{equation}%
Thus, the transformation of the temperature (\ref{transf-T}) accompanies the
action (\ref{transf-S}) of the Freudenthal map onto the near-extremal
entropy (\ref{eq:newnonen}) as
\begin{equation}
S_{NE}(\mathcal{Q};T)\longrightarrow \mathfrak{F}_{T+\delta T}\left(
S_{NE}\left( \mathcal{Q};T+\delta T\right) \right) .  \label{comb-transf}
\end{equation}%
Consequently, the condition of invariance of the near-extremal entropy under
the combined action of Freudenthal duality and temperature transformation (%
\ref{transf-T}) reads%
\begin{equation}
\mathfrak{F}_{T+\delta T}\left( S_{NE}\left( \mathcal{Q};T+\delta T\right)
\right) =S_{NE}(\mathcal{Q};T),  \label{condgf}
\end{equation}%
which should be solved in terms of $\delta T$ (\ref{delta-T}). Physically
speaking, we are aiming at finding two near-extremal black holes, with small
temperatures $T$ and $T+\delta T$, such that they have the same entropy and
their charges are related by the Freudenthal map defined by (\ref{F_T})
evaluated at $T+\delta T$:%
\begin{eqnarray}
\mathcal{Q}\longrightarrow \mathfrak{F}_{T+\delta T}\left( \mathcal{Q}%
\right) \equiv \hat{\mathcal{Q}}\left( \mathcal{Q};T+\delta T\right) :&=
&\Omega \frac{\partial S_{NE}(\mathcal{Q};T+\delta T)}{\partial \mathcal{Q}}
\notag \\
&=&\Omega \frac{\partial S_{0}\left( \mathcal{Q}\right) }{\partial \mathcal{Q%
}}+\frac{3\alpha }{2}\left( T+\delta T\right) \sqrt{S_{0}\left( \mathcal{Q}%
\right) }\Omega \frac{\partial S_{0}\left( \mathcal{Q}\right) }{\partial
\mathcal{Q}}  \notag \\
&=&\left( 1+\frac{3\alpha }{2}\left( T+\delta T\right) \sqrt{S_{0}\left(
\mathcal{Q}\right) }\right) \mathfrak{F}_{0}\left( \mathcal{Q}\right) .
\label{F_(T+deltaT)}
\end{eqnarray}%
Correspondingly, the action of the Freudenthal map $\mathfrak{F}_{T+\delta
T} $ onto the near-extremal black hole entropy $S_{NE}(\mathcal{Q};T)$ (\ref%
{eq:newnonen}) is given by (\ref{comb-transf}), as%
\begin{eqnarray}  \label{new-transf-S}
\mathfrak{F}_{T+\delta T}\left( S_{NE}\left( \mathcal{Q};T+\delta T\right)
\right) &=&S_{NE}(\mathfrak{F}_{T+\delta T}\left( \mathcal{Q}\right)
;T+\delta T)  \notag \\
&=&S_{0}\left( \mathfrak{F}_{T+\delta T}\left( \mathcal{Q}\right) \right)
+\alpha \left( T+\delta T\right) S_{0}^{3/2}\left( \mathfrak{F}_{T+\delta
T}\left( \mathcal{Q}\right) \right)  \notag \\
&=&\left( 1+\frac{3\alpha }{2}\left( T+\delta T\right) \sqrt{S_{0}\left(
\mathcal{Q}\right) }\right) ^{2}S_{0}\left( \mathcal{Q}\right)  \notag \\
&&+\alpha \left( T+\delta T\right) \left( 1+\frac{3\alpha }{2}\left(
T+\delta T\right) \sqrt{S_{0}\left( \mathcal{Q}\right) }\right)
^{3}S_{0}^{3/2}\left( \mathcal{Q}\right)  \notag \\
&=&\left( 1+\frac{3\alpha }{2}\left( T+\delta T\right) \sqrt{S_{0}\left(
\mathcal{Q}\right) }\right) ^{2} \Bigg[ 1+\alpha \left( T+\delta T\right) %
\Big( 1+  \notag \\
&& \frac{3\alpha }{2}\left( T+\delta T\right) \sqrt{S_{0}\left( \mathcal{Q}%
\right) }\Big) \sqrt{S_{0}\left( \mathcal{Q}\right) }\Bigg] S_{0}\left(
\mathcal{Q}\right) ,
\end{eqnarray}%
Therefore, by virtue of (\ref{new-transf-S}), the condition of invariance (%
\ref{condgf}) of the near-extremal entropy yields the following algebraic,
inhomogeneous equation of degree four in $\delta T$ as
\begin{equation}
\left( \delta T\right) ^{4}+a_{3}\left( \delta T\right) ^{3}+a_{2}\left(
\delta T\right) ^{2}+a_{1}\delta T+a_{0}=0,  \label{eq:quartic}
\end{equation}
with
\begin{equation}
\begin{split}
a_{3}& :=2\left( \frac{1}{A}+2T\right) ; \\
a_{2}& :=2\left( \frac{1}{A^{2}}+\frac{3T}{A}+3T^{2}\right) ; \\
a_{1}& :=4\left( \frac{8}{27A^{3}}+\frac{T}{A^{2}}+\frac{3T^{2}}{2A}%
+T^{3}\right) ; \\
a_{0}& :=T\left( \frac{8}{9A^{3}}+\frac{2T}{A^{2}}+\frac{2T^{2}}{A}%
+T^{3}\right) ,
\end{split}
\label{eq:alla}
\end{equation}%
and
\begin{equation}
A:=\alpha \sqrt{S_{0}\left( \mathcal{Q}\right) }.  \label{A}
\end{equation}%
The physically consistent (namely, real) solutions to the quartic
inhomogeneous Eq. (\ref{eq:quartic}) will determine the analytical
functional form of (\ref{delta-T}). At best, Eq. (\ref{eq:quartic}) admits
four real, analytical solutions, e.g. given by (see e.g. \cite%
{abramowitz65HMF}, p. 17)
\begin{equation}
\begin{split}
\delta T_{1}& :=-\frac{a_{3}}{4}+\frac{R}{2}+\frac{D}{2}; \\
\delta T_{2}& :=-\frac{a_{3}}{4}+\frac{R}{2}-\frac{D}{2}; \\
\delta T_{3}& :=-\frac{a_{3}}{4}-\frac{R}{2}+\frac{E}{2}; \\
\delta T_{4}& :=-\frac{a_{3}}{4}-\frac{R}{2}-\frac{E}{2},
\end{split}
\label{eq:sols}
\end{equation}%
where
\begin{eqnarray}
R &:&=\sqrt{\frac{a_{3}^{2}}{4}-a_{2}+y};  \label{eq:RDE} \\
D &:&=\left\{
\begin{array}{l}
\sqrt{\frac{3}{4}a_{3}^{2}-R^{2}-2a_{2}+\frac{1}{4R}\left(
4a_{2}a_{3}-8a_{1}-a_{3}^{3}\right) }~\text{for}~R\neq 0, \\
\sqrt{\frac{3}{4}a_{3}^{2}-2a_{2}+2\sqrt{y^{2}-4a_{0}}}~\text{for~}R=0;%
\end{array}%
\right.  \label{eq:RDE2} \\
E &:&=\left\{
\begin{array}{l}
\sqrt{\frac{3}{4}a_{3}^{2}-R^{2}-2a_{2}-\frac{1}{4R}\left(
4a_{2}a_{3}-8a_{1}-a_{3}^{3}\right) }~\text{for}~R\neq 0, \\
\sqrt{\frac{3}{4}a_{3}^{2}-2a_{2}-2\sqrt{y^{2}-4a_{0}}}~\text{for~}R=0,%
\end{array}%
\right.  \label{eq:RDE3}
\end{eqnarray}%
and $y$ is a real root of the algebraic inhomogeneous equation of degree
three
\begin{equation}
y^{3}-a_{2}y^{2}+\left( a_{1}a_{3}-4a_{0}\right) y+\left(
4a_{0}a_{2}-a_{1}^{2}-a_{0}a_{3}^{2}\right) =0.  \label{eq:yeqn}
\end{equation}%
The four solutions (\ref{eq:sols}) will generally express $\delta T$ as a
(real, analytical) function of $T$ and $A$, and thus of $T$ and $S_{0}\left(%
\mathcal{Q}\right)$, as given by (\ref{delta-T}). Therefore, as we said,
four different real, analytical functions $f_{i}\left( T,A\right) :=\delta
T_{i}$ ($i=1,...,4$) are at best possible. Eqs. (\ref{eq:sols}) defines four
possible sets of transformations of the temperature $T$ of an asymptotically
flat, static, spherically symmetric near-extremal black hole with e.m.
charge $\mathcal{Q}$, mapping it to its \textit{Freudenthal dual}
(asymptotically flat, static, spherically symmetric) near extremal black
hole with the e.m. charges $\mathfrak{F}_{T+f_{i}\left( T,A\right) }\left(
\mathcal{Q}\right)$ and the temperature $T+f_{i}\left( T,A\right)$, i.e.,
\begin{equation}
\left\{
\begin{array}{l}
T\longrightarrow T+f_{i}\left( T,A\right) ; \\
\\
\mathcal{Q}\longrightarrow \mathfrak{F}_{T+f_{i}\left( T,A\right) }\left(
\mathcal{Q}\right) ,%
\end{array}%
\right.  \label{eq:possibilities}
\end{equation}%
such that
\begin{equation}
\mathfrak{F}_{T+f_{i}\left( T,A\right) }\left( S_{NE}\left( \mathcal{Q}%
;T+f_{i}\left( T,A\right) \right) \right) =S_{NE}(\mathcal{Q};T),~\forall i,
\label{sameS}
\end{equation}%
where we understood that%
\begin{equation}
\mathfrak{F}_{T+f_{i}\left( T,A\right) }\left( S_{NE}\left( \mathcal{Q}%
;T+f_{i}\left( T,A\right) \right) \right) =S_{NE}\left( \mathfrak{F}%
_{T+f_{i}\left( T,A\right) }\left( \mathcal{Q}\right) ;T+f_{i}\left(
T,A\right) \right) .
\end{equation}

Let us analyse the solutions (\ref{eq:sols}) more carefully. We are looking
for a physical solution $\delta T$ such that, for a given (small)
temperature $T$, the temperature $T+\delta T$ is always positive (and
small). As $A$ (\ref{A}) and $T$ are both positive and real, so are the
coefficients $a_{0},a_{1},a_{2},a_{3}$ of the quartic equation (\ref%
{eq:quartic}), as defined in (\ref{eq:alla}). By applying \textit{Descartes'
rule of signs} (see App. \ref{app:sturm}), one can realize that Eq. (\ref%
{eq:quartic}) does \textit{not have any positive real root}, but it rather
can have four, two or zero negative roots. Thus, in order to select a
physically meaningful solution, we need to show that Eq. (\ref{eq:quartic}),
with coefficients (\ref{eq:alla}), has \textit{at least one} real negative
root $\delta T$, such that%
\begin{equation}
\delta T\in \left( -T,0\right) \Leftrightarrow 0<\left\vert \delta
T\right\vert =-\delta T<T,  \label{ineq1}
\end{equation}%
otherwise, the Freudenthal dual near-extremal black hole will not have a
real and positive temperature $T+\delta T=T-\left\vert \delta T\right\vert $.

\subsection{\label{Uniqueness}Uniqueness of $\protect\delta T$}

In this subsection, without explicitly solving (\ref{eq:quartic}), we will
prove and verify that such a quartic equation does always have a unique root
$\delta T$ satisfying (\ref{ineq1}), namely does always have an \textit{%
unique physically sensible solution} $\delta T$.

In order to achieve such a result, we will invoke \textit{Sturm's Theorem},
which is recalled in App. \ref{app:sturm}. To get started with the root
analysis, we write down the so-called \textit{Sturm's sequence} for the
quartic equation (\ref{eq:quartic}) with coefficients (\ref{eq:alla}) : by
denoting $\delta T\equiv x$, such a finite sequence is made of five
polynomials $p_{I}(x)$ with $I=0,1,...,4$, such that

\begin{enumerate}
\item $p_{0}(x)=0$ is nothing but Eq. (\ref{eq:quartic}) (with coefficients (%
\ref{eq:alla}) made explicit\footnote{%
For simplicity's sake, we set $G_{4}=1$, which implies that $\alpha =4\sqrt{%
\pi }$.});

\item $p_{1}(x)={\frac{dp_{0}(x)}{dx};}$

\item $p_{2}(x)=-$Rem$\left( p_{0}(x),p_{1}(x)\right) ;$

\item $p_{3}(x)=-$Rem$\left( p_{1}(x),p_{2}(x)\right) ;$

\item $p_{4}=-$Rem$\left( p_{2}(x),p_{3}(x)\right) ,$
\end{enumerate}

where, for non zero polynomials $a(x)$ and $b(x)$, Rem$(a(x),b(x))$ denotes
the remainder of the Euclidean division of $a(x)$ by $b(x)$. The degree in $%
x $ of the (generally inhomogeneous) polynomials $p_{I}(x)$ is $4-I$, so
actually $p_{4}$ does \textit{not} depend on $x$ : $p_{4}\neq p_{4}(x)$.
Further details can be found in App. \ref{app:sturm}. Explicitly, one obtains%
\footnote{%
Note that $p_{4}$ is always negative.}
\begin{eqnarray}
p_{0}(\delta T) &=&(\delta T+T)^{4}+{\frac{(\delta T+T)^{3}}{2\sqrt{\pi S_{0}%
}}}+{\frac{(\delta T+T)^{2}}{8\pi S_{0}}}+{\frac{4\delta T+3T}{216\pi ^{%
\frac{3}{2}}S_{0}^{\frac{3}{2}}};} \\
p_{1}(\delta T) &=&4(\delta T+T)^{3}+{\frac{3(\delta T+T)^{2}}{2\sqrt{\pi
S_{0}}}}+{\frac{\delta T+T}{4\pi S_{0}}}+{\frac{1}{54\pi ^{\frac{3}{2}%
}S_{0}^{\frac{3}{2}}};} \\
p_{2}(\delta T) &=&{\frac{-54\pi S_{0}(\delta T+T)^{2}-\sqrt{\pi S_{0}}%
(21\delta T+5T)+2}{3456\pi ^{2}S_{0}^{2}};} \\
p_{3}(\delta T) &=&-{\frac{2\left( 2+144\pi S_{0}T(\delta T+T)+\sqrt{\pi
S_{0}}(51\delta T+49T)\right) }{243\pi ^{\frac{3}{2}}S_{0}^{\frac{3}{2}}};}
\\
p_{4} &=&-{\frac{(1+4\sqrt{\pi S_{0}}T)(44+289\sqrt{\pi S_{0}}T+512\pi
S_{0}T^{2})}{192\pi ^{2}S_{0}^{2}(17+48\sqrt{\pi S_{0}}T)^{2}}.}
\end{eqnarray}

Next, the application of \textit{Descartes' rule of signs} (discussed in
App. \ref{app:descarte}) yields that no positive real roots exist for the
quartic equation (\ref{eq:quartic}); thus, the unique physically sensible
domain for a root $\delta T$ is specified by (\ref{ineq1}), namely $\delta
T\in \left( -T,0\right) $. Thus, we need to compute the signs of the limits
of $p_{I}\left( \delta T\right) $ (with $I=0,1,2,3$) for $\delta
T\rightarrow 0^{-}$ and $\delta T\rightarrow -T^{+}$. For what concerns $%
\lim_{\delta T\rightarrow 0^{-}}p_{I}\left( \delta T\right) $ and their
signs, one obtains\footnote{%
Recall that $T$ and $S_{0}\equiv S_{0}(\mathcal{Q})$ are always both real
and positive.}
\begin{eqnarray}
\lim_{\delta T\rightarrow 0^{-}}p_{0}(\delta T) &=&{\frac{1}{72}}T\left( {%
\frac{1}{\pi ^{\frac{3}{2}}S_{0}^{\frac{3}{2}}}}+{\frac{9T}{\pi S_{0}}}+{%
\frac{36T^{2}}{\sqrt{\pi S_{0}}}}+72T^{3}\right) >0; \\
\lim_{\delta T\rightarrow 0^{-}}p_{1}(\delta T) &=&{\frac{1}{54\pi ^{\frac{3%
}{2}}S_{0}^{\frac{3}{2}}}}+{\frac{T}{4\pi S_{0}}}+{\frac{3T^{2}}{2\sqrt{\pi
S_{0}}}}+4T^{3}>0; \\
\lim_{\delta T\rightarrow 0^{-}}p_{2}(\delta T) &=&{\frac{2-5\sqrt{\pi S_{0}}%
T-54\pi S_{0}T^{2}}{3456\pi ^{2}S_{0}^{2}}\gtreqless 0\Leftrightarrow }%
T\lesseqgtr C_{1};  \label{eq:crit1} \\
\lim_{\delta T\rightarrow 0^{-}}p_{3}(\delta T) &=&-{\frac{2(2+49\sqrt{\pi
S_{0}}T+144\pi S_{0}T^{2})}{243\pi ^{\frac{3}{2}}S_{0}^{\frac{3}{2}}}<0,}
\end{eqnarray}%
with%
\begin{equation}
C_{1}:={\frac{-5+\sqrt{457}}{108\sqrt{\pi S_{0}}}}\approx {\frac{0.08555}{%
\sqrt{S_{0}}}.}
\end{equation}%
Thus, recalling that $p_{4}$ is always negative, for the signs of $%
\lim_{\delta T\rightarrow 0^{-}}p_{I}(\delta T)$ (for $I=0,1,..,4$) we
obtain the sequence%
\begin{equation}
\{+,+,\pm ,-,-\}~\text{for~}T\lessgtr C_{1}.  \label{seq-signs-1}
\end{equation}%
Regardless of whether $T\lessgtr C_{1}$, there is always only one change in
the sequence of signs (\ref{seq-signs-1}). By denoting with $\sigma (\alpha
) $ the \textit{number of sign changes function} evaluated at the point $%
\alpha $ (cfr. App. \ref{app:sturm}), we obtain that, regardless of whether $%
T\lessgtr C_{1}$,
\begin{equation}
\sigma (0)=1.
\end{equation}

On the other hand, for what concerns $\lim_{\delta T\rightarrow
-T^{+}}p_{I}\left( \delta T\right) $ and their signs, one obtains
\begin{eqnarray}
\lim_{\delta T\rightarrow -T^{+}}p_{0}(\delta T) &=&-{\frac{T}{216\pi ^{%
\frac{3}{2}}S_{0}^{\frac{3}{2}}}<0;} \\
\lim_{\delta T\rightarrow -T^{+}}p_{1}(\delta T) &=&{\frac{1}{54\pi ^{\frac{3%
}{2}}S_{0}^{\frac{3}{2}}}>0;} \\
\lim_{\delta T\rightarrow -T^{+}}p_{2}(\delta T) &=&{\frac{1+8\sqrt{\pi S_{0}%
}T}{1728\pi ^{2}S_{0}^{2}}>0;} \\
\lim_{\delta T\rightarrow -T^{+}}p_{3}(\delta T) &=&{\frac{4(-1+\sqrt{\pi
S_{0}}T)}{243\pi ^{\frac{3}{2}}S_{0}^{\frac{3}{2}}}\gtreqless
0\Leftrightarrow }T\gtreqless C_{2},  \label{eq:crit2}
\end{eqnarray}%
with%
\begin{equation}
C_{2}:={\frac{1}{\sqrt{\pi S_{0}}}}\approx {\frac{0.56419}{\sqrt{S_{0}}}.}
\end{equation}%
Thus, recalling again that $p_{4}$ is always negative, for the signs of $%
\lim_{\delta T\rightarrow -T^{+}}p_{I}(\delta T)$ (for $I=0,1,..,4$) we
obtain the sequence%
\begin{equation}
\{-,+,+,\pm ,-\}~\text{for~}T\gtrless C_{2}.  \label{seq-signs-2}
\end{equation}%
Regardless whether $T\gtrless C_{2}$, there are always two changes in the
sequence of signs (\ref{seq-signs-2}); thus, regardless whether $T\gtrless
C_{2}$,
\begin{equation}
\sigma (-T)=2.
\end{equation}

Thus, by virtue of Sturm's Theorem, discussed in App. \ref{app:sturm}, since%
\begin{equation}
\sigma (-T)-\sigma (0)=1,
\end{equation}%
it follows that : $\forall T$ and $S_{0}\in \mathbb{R}_{0}^{+}$ in the
quartic equation (\ref{eq:quartic}), $\exists !$ $\delta T\in \mathbb{R}%
_{0}^{-}:T+\delta T\in \mathbb{R}_{0}^{+}$. Since $T$ is assumed to be small
(from \textit{near-extremality} of black holes under consideration) and $%
\delta T$ satisfies (\ref{ineq1}), $T+\delta T$ is also small. In other
words, the near-extremal Freudenthal transformation (\ref{eq:possibilities})
for $\delta T\equiv f\left( T,A\right) $\emph{\ }satisfying (\ref{sameS})
\textit{uniquely maps two near-extremal black holes with the same entropy
but different (small) temperatures.}

\subsection{\label{Precision}$\protect\delta T=\protect\delta T_{3}$}

So far, we have shown that the near-extremal Freudenthal duality
transformation (\ref{eq:possibilities})\emph{\ }satisfying (\ref{sameS})
always admits a physically sensible solution, which is also unique. In fact,
by exploiting Sturm's Theorem, we have been able to show that (\ref%
{eq:quartic}) has precisely one real root, ranging as $-T<\delta T<0$. We
will now single out such a sensible solution among the four possible ones
determined by formula (\ref{eq:sols}).

We start and recall again that both $T$ and $S_{0}\equiv S_{0}(\mathcal{Q})$
are always real and positive, and this implies that all coefficients $%
a_{0},\,a_{1},\,a_{2},\,a_{3}$, defined by (\ref{eq:alla}) are real and
positive. Moreover, as proved in App. \ref{app:sturm}, the cubic
inhomogeneous Eq. (\ref{eq:yeqn}) always admits a unique real root, which is
positive. By recalling (\ref{eq:alla}) as well as the definition of $R$
given by (\ref{eq:RDE}), we observe that
\begin{equation}
R^{2}-y={\frac{a_{3}^{2}}{4}}-a_{2}=-{\frac{1+8\sqrt{\pi S_{0}}T+32\pi
S_{0}T^{2}}{16\pi S_{0}}<0,}
\end{equation}%
and $R\neq 0$ for any value of $T$ and $S_{0}$. Because%
\begin{equation}
R=0\overset{\text{(\ref{eq:RDE})}}{\Leftrightarrow }y=a_{2}-\frac{a_{3}^{2}}{%
4},  \label{yeqn2}
\end{equation}%
but the algebraic system made by (\ref{eq:yeqn}) and (\ref{yeqn2}) has no
solution\footnote{%
Explicitly, for $y=a_{2}-{\frac{a_{3}^{2}}{4}}$, it holds that $%
y^{3}-a_{2}y^{2}+\left( a_{1}a_{3}-4a_{0}\right) y+\left(
4a_{0}a_{2}-a_{1}^{2}-a_{0}a_{3}^{2}\right) =-{\frac{25}{2985984\pi
^{3}S_{0}^{3}}}$.}.

Let us now introduce a trick to show that $R$ is always real and positive.
Under the reparametrisation $y^{\prime }:=y-a_{2}+{\frac{a_{3}^{2}}{4}}$,
Eq. (\ref{eq:yeqn}) becomes
\begin{equation}
y^{\prime 3}+\left( 2a_{2}-{\frac{3a_{3}^{2}}{4}}\right) y^{\prime 2}+\left(
a_{2}^{2}-4a_{0}+a_{1}a_{3}-a_{2}a_{3}^{2}+{\frac{3a_{3}^{4}}{16}}\right)
y^{\prime }-{\frac{1}{64}}\left( 8a_{1}-4a_{2}a_{3}+a_{3}^{3}\right) ^{2}=0,
\end{equation}%
which simplifies into
\begin{equation}
y^{\prime 3}+{\frac{1}{16\pi S_{0}}}y^{\prime 2}+{\frac{\left( 37+128\sqrt{%
\pi S_{0}}T\right) }{6912\pi ^{2}S_{0}^{2}}}y^{\prime }-{\frac{25}{%
2985984\pi ^{3}S_{0}^{3}}}=0;  \label{eq:y'eqn}
\end{equation}%
By applying \textit{Descartes' rule of signs} (cf. App. \ref{app:descarte}),
one can establish that Eq. (\ref{eq:y'eqn}) always admits only one real root
$y^{\prime }$, which is positive. Since we already mentioned that Eq. (\ref%
{eq:yeqn}) always admits only one real root $y$, which is positive, one
obtains that%
\begin{equation}
\left.
\begin{array}{r}
y^{\prime }:=y-a_{2}+{\frac{a_{3}^{2}}{4}>0;} \\
y>0;%
\end{array}%
\right\} \Rightarrow y>a_{2}-{\frac{a_{3}^{2}}{4}>0}\Rightarrow R\in \mathbb{%
R}_{0}^{+}.  \label{ress}
\end{equation}

Next, moving onto the analysis of the term $D$ defined in the formula (\ref%
{eq:RDE2}), for $R\neq 0$,
\begin{equation*}
D:=\sqrt{\frac{3}{4}a_{3}^{2}-R^{2}-2a_{2}+\frac{1}{4R}\left(
4a_{2}a_{3}-8a_{1}-a_{3}^{3}\right) },
\end{equation*}
one should firstly note that the Eq. (\ref{eq:alla}) yield
\begin{eqnarray}
{\frac{3}{4}}a_{3}^{2}-2a_{2} &=&-{\frac{1}{16\pi S_{0}};} \\
4a_{2}a_{3}-8a_{1}-a_{3}^{3} &=&-{\frac{5}{216\pi ^{\frac{3}{2}}S_{0}^{\frac{%
3}{2}}}},
\end{eqnarray}%
implying
\begin{equation}
D^{2}=\frac{3}{4}a_{3}^{2}-R^{2}-2a_{2}+\frac{1}{4R}\left(
4a_{2}a_{3}-8a_{1}-a_{3}^{3}\right) <0,  \label{D-im}
\end{equation}%
Where the definition (\ref{eq:RDE2}) and the result (\ref{ress}) have been
used. Thus, $D\in i\mathbb{R}_{0}$, and the solutions $\delta T_{1}$ and $%
\delta T_{2}$ in the formula (\ref{eq:sols}) acquire a non-vanishing
imaginary part. Hence they are to be discarded as they are \textit{unphysical%
}. In general, the quartic inhomogeneous Eq. (\ref{eq:quartic}) always
admits two complex conjugate roots $\delta T_{1}$ and $\delta T_{2}$, as
defined by formula (\ref{eq:sols}).

As discussed above, \textit{Descartes' rule of signs} (cf. App. \ref%
{app:descarte}) yields that the quartic Eq. (\ref{eq:quartic}) has no
positive real roots, but rather it can have four, two or zero negative
roots. The result obtained on the complex nature of $\delta T_{1}$ and $%
\delta T_{2}$ allows us to discard the case of four negative real roots.
Whereas the proof given above (through the use of \textit{Sturm's Theorem};
cf. App. \ref{app:sturm}) that exactly one real negative root in the
interval $(-T,0)$ is admitted by Eq. (\ref{eq:quartic}) allows us to discard
the case of zero negative real roots. Thus, we can conclude that, for any
real positive values of $T$ and $S_{0}$, Eq. (\ref{eq:quartic}) admits two
complex conjugate solutions $\delta T_{1}$ and $\delta T_{2}$ defined by (%
\ref{eq:sols}), as well as two real negative roots $\delta T_{3}$ and $%
\delta T_{4}$ defined by (\ref{eq:sols}). Among the two negative real roots,
only one lies in the physically admissible range $\left( -T,0\right) $, and
thus satisfies (\ref{ineq1}). At this point, we have to establish which one
between $\delta T_{3}$ and $\delta T_{4}$, defined by (\ref{eq:sols}) is a
physically sensible real negative root. We observe that in the same way we
proved (\ref{D-im}), we can also prove that%
\begin{equation}
E^{2}\overset{\text{(\ref{eq:alla})}}{=}\frac{3}{4}a_{3}^{2}-R^{2}-2a_{2}-%
\frac{1}{4R}\left( 4a_{2}a_{3}-8a_{1}-a_{3}^{3}\right) >0,  \label{E-real}
\end{equation}%
thus yielding $E\in \mathbb{R}_{0}^{+}$. Thus, from the fourth of (\ref%
{eq:sols}) it follows that%
\begin{equation}
T+\delta T_{4}=T-\frac{a_{3}}{2}-\frac{1}{2}\left( R+E\right) \overset{\text{%
(\ref{eq:alla})}}{=}-T-\frac{1}{A}-\frac{1}{2}\left( R+E\right) <0,
\end{equation}%
where in the last step we have used the fact that $T$, $A$, $R$ and also $E$
are all real and positive. Hence, the negative real root $\delta T_{4}$
defined by (\ref{eq:sols}) is unphysical, and it must be discarded. By
recalling the outcome of the root analysis based on Sturm's Theorem done
above, we can conclude that the negative real root $\delta T_{3}$ defined by
(\ref{eq:sols}) is the physically admissible and sensible one, since it
satisfies the inequality (\ref{ineq1}), i.e. it holds that $-T<\delta
T_{3}<0 $.

\begin{figure}[tbph]
\begin{subfigure}{0.48\textwidth}
		\centering
		\includegraphics[width=\textwidth]{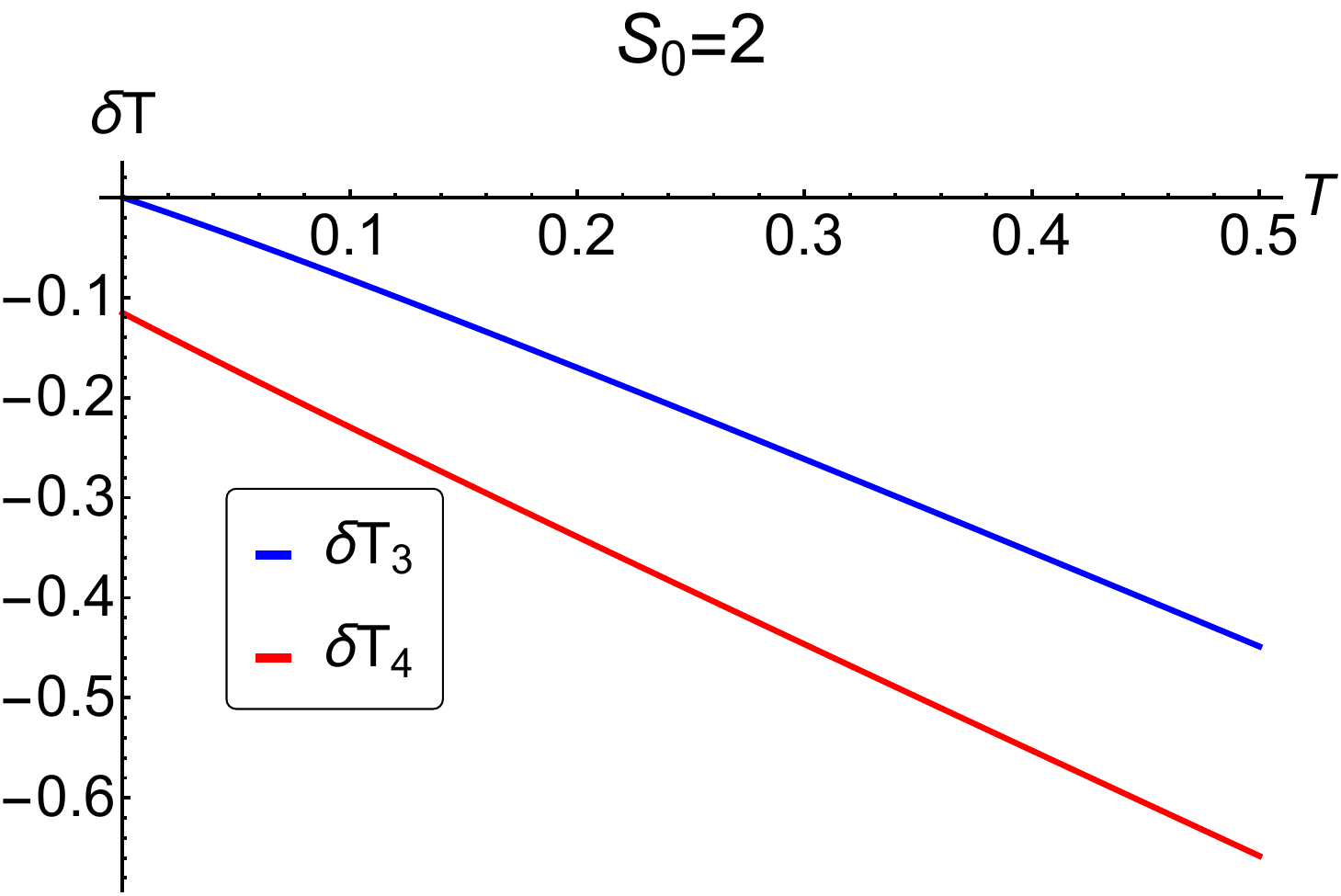}
		\caption{Fixed Entropy}
	\end{subfigure}\hfill
\begin{subfigure}{0.48\textwidth}
		\centering
		\includegraphics[width=\textwidth]{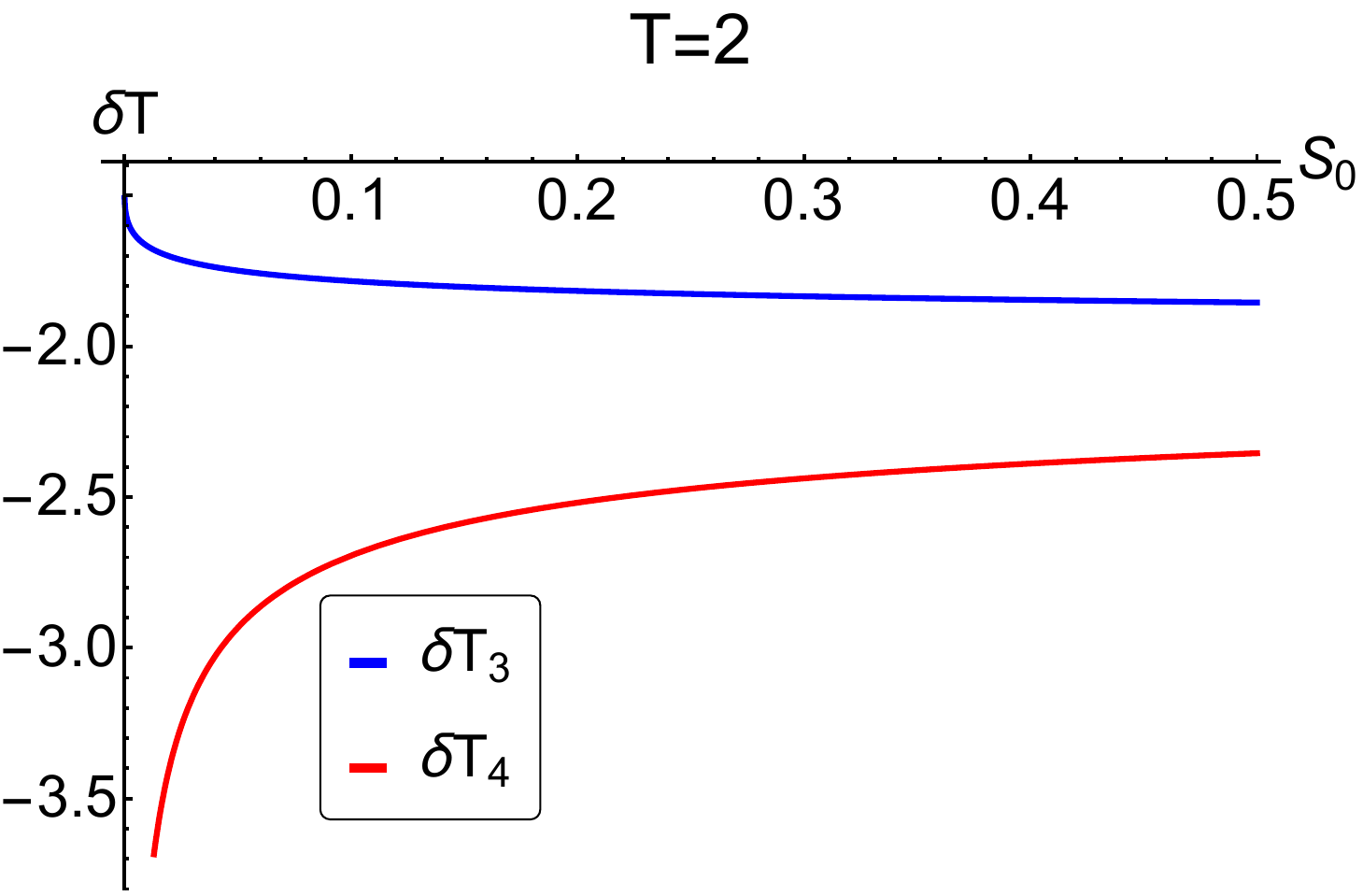}
		\caption{Fixed temperature}
	\end{subfigure}
\caption{Showing the two negative roots of (\protect\ref{eq:quartic})}
\label{fig:tdt}
\end{figure}

\begin{figure}[tbph]
\begin{subfigure}{0.48\textwidth}
		\centering
		\includegraphics[width=\textwidth]{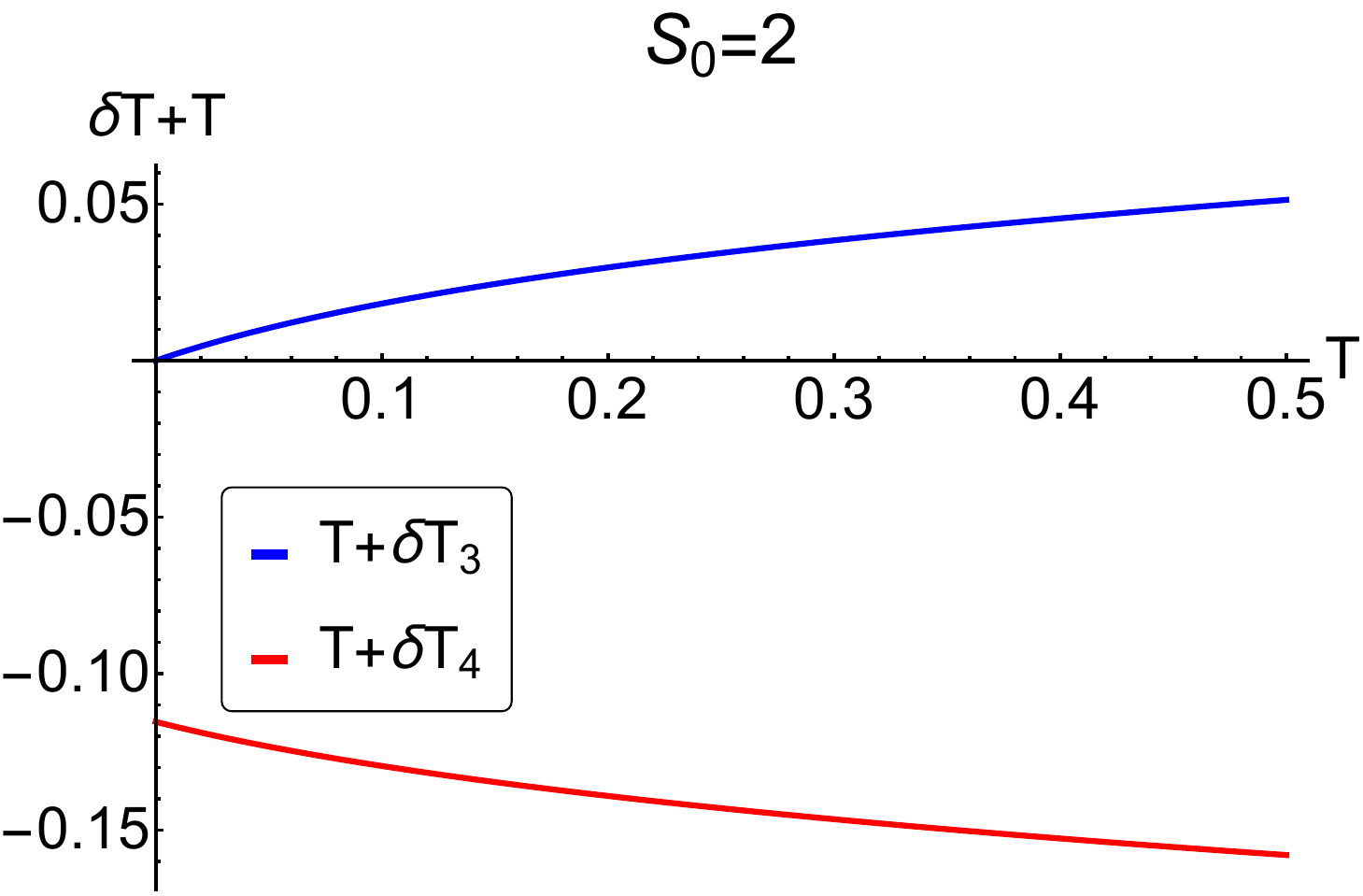}
		\caption{Fixed Entropy}
	\end{subfigure}\hfill
\begin{subfigure}{0.48\textwidth}
		\centering
		\includegraphics[width=\textwidth]{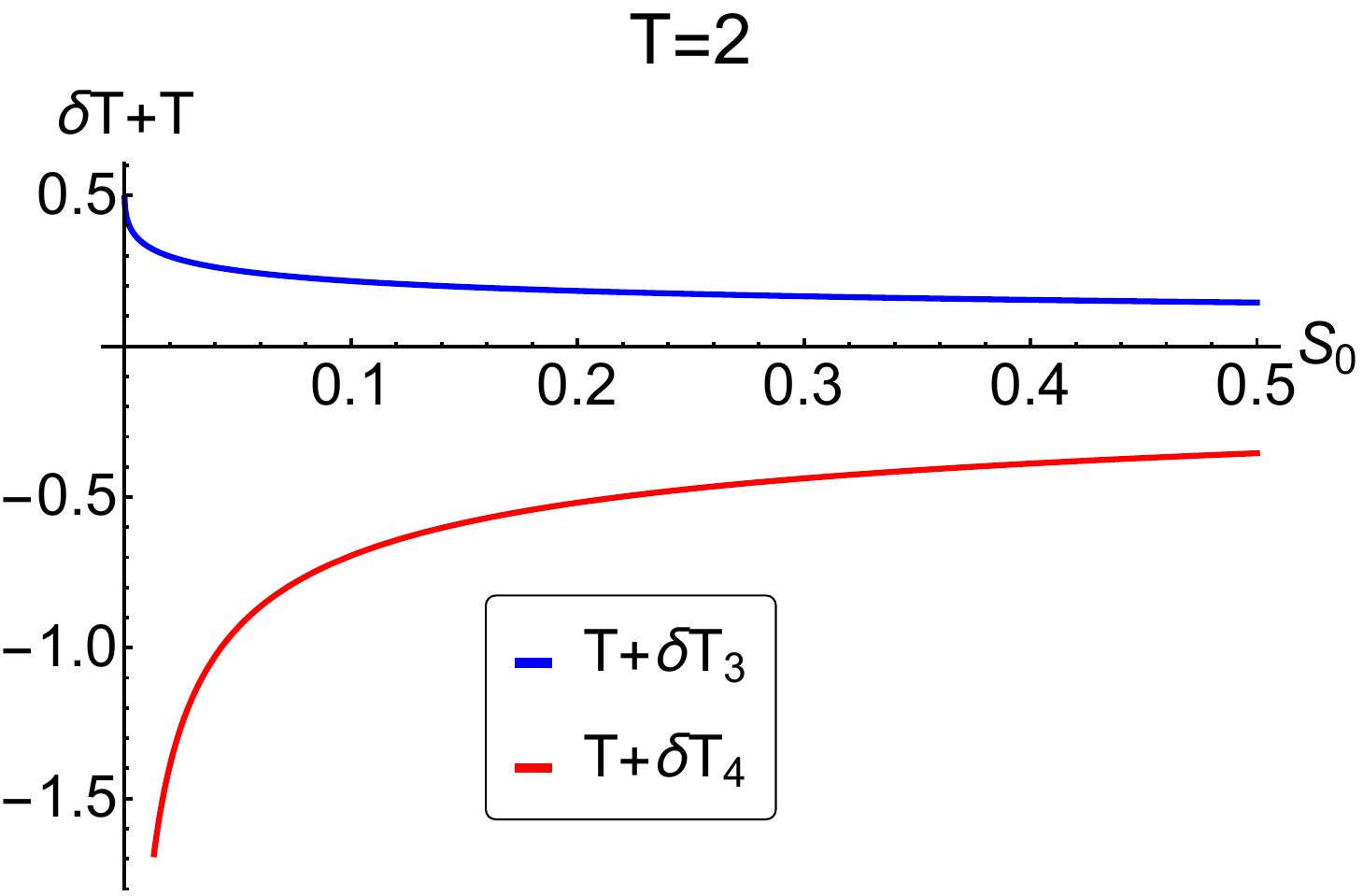}
		\caption{Fixed temperature}
	\end{subfigure}
\caption{Showing the Freudenthal dual temperature $T+\protect\delta T$
corresponding to the two negative real roots $\protect\delta T_{3}$ and $%
\protect\delta T_{4}$ of (\protect\ref{eq:quartic})}
\label{fig:tdtwT}
\end{figure}

The validity of $\delta T_{3}$ as the physically admissible real roots of
the quartic inhomogeneous equation (\ref{eq:quartic}) can also be confirmed
through numerical calculation, as shown in Fig. \ref{fig:tdt} and in Fig. %
\ref{fig:tdtwT}.
Fig. \ref{fig:tdt} shows that both $\delta T_{3}$ and $\delta T_{4}$ are
real and negative, as discussed. On the other hand, the Freudenthal dual
black hole temperature $T+\delta T$ is positive only for $\delta T_{3}$, as
shown in Fig. \ref{fig:tdtwT}. In each figure, we plot two cases, namely
when considering fixed entropy $S_{0}$ and when considering fixed
temperature $T$.

Thus, we have finally determined the near-extremal Freudenthal
transformation (\ref{eq:possibilities}) for $\delta T=\delta T_{3}\equiv
f_{3}\left( T,A\right) $\emph{\ }which satisfies (\ref{sameS}). This
non-linear transformation of the e.m. charges $\mathcal{Q}$ and of the
(small) temperature $T$ of a given near-extremal (asymptotically flat,
static, spherically symmetric) black hole maps it to another near-extremal
(asymptotically flat, static, spherically symmetric) with Freudenthal dual
e.m. charges $\mathfrak{F}_{T+\delta T_{3}}\left( \mathcal{Q}\right) $ and
Freudenthal dual temperature $T+\delta T_{3}$, but with the same entropy,
namely, it holds%
\begin{equation}
\mathfrak{F}_{T+\delta T_{3}}\left( S_{NE}\left( \mathcal{Q};T+\delta
T_{3}\right) \right) =S_{NE}(\mathcal{Q};T).  \label{jj}
\end{equation}%
In other words, we have precisely selected $i=3$ in Eqs. (\ref%
{eq:possibilities}) and (\ref{sameS}), disregarding the other three values $%
i=1,2$ and $4$ (within the labelling defined by formula (\ref{eq:sols})).

\textbf{Remark}

It is here worth remarking that the \textit{near-extremal (on-shell)
Freudenthal duality} $\mathfrak{F}$, defined by (\ref{eq:possibilities}) and
(\ref{sameS}) with $i=3$ (cf. (\ref{jj})), \textit{cannot} determine a
transition to and from the extremal limit :
\begin{equation}
\text{near-extremal}\overset{\mathfrak{F}}{%
\begin{array}{c}
\nRightarrow \\
\nLeftarrow%
\end{array}%
}\text{extremal.}
\end{equation}

\textbf{Proof}

\begin{description}
\item[$\nLeftarrow $] Since the $\delta T$ pertaining to $\mathfrak{F}$
(i.e., uniquely satisfying (\ref{eq:quartic}) and (\ref{ineq1})) is \textit{%
negative}, $-T<\delta T<0$, it follows that $\mathfrak{F}$ \textit{cannot}
be used to make a transition from the extremal state ($T=0$) to a
near-extremal one ($T>0$), and keep the Bekenstein-Hawking entropy/area of
the horizon fixed. In other words, the following transformation :%
\begin{equation}
\left\{
\begin{array}{l}
\mathcal{Q}\longrightarrow \mathfrak{F}_{\delta T}\left( \mathcal{Q}\right)
\equiv \hat{\mathcal{Q}}\left( \mathcal{Q};\delta T\right) :=\Omega \frac{%
\partial S_{NE}(\mathcal{Q};\delta T)}{\partial \mathcal{Q}}; \\
~ \\
T=0\longrightarrow T=\delta T,%
\end{array}%
\right. ~\text{such~that~}S_{0}\left( \mathcal{Q}\right) =S_{\delta T}\left(
\hat{\mathcal{Q}}\left( \mathcal{Q};\delta T\right) \right) ,
\end{equation}%
\textit{cannot} occur, simply because $\delta T<0$.

\item[$\nRightarrow $] On the other hand, it can be proved \textit{per
absurdum} that the same $\delta T$ cannot reach the limit value of $-T$. In
fact, if we set $\delta T=-T$, one could perform a $\mathfrak{F}$%
-transformation from a near-extremal state ($T\neq 0$) to the extremal state
($T=0$), and keep the black hole entropy fixed,but this is impossible. In
other words, by recalling the definition (\ref{eq:s0def}), the following
transformation :%
\begin{equation}
\left\{
\begin{array}{l}
\mathcal{Q}\longrightarrow \mathfrak{F}_{0}\left( \mathcal{Q}\right)
:=\Omega \frac{\partial S_{0}(\mathcal{Q})}{\partial \mathcal{Q}}; \\
~ \\
T\longrightarrow T+\delta T=0,%
\end{array}%
\right. ~\text{such~that~}S_{NE}\left( \mathcal{Q};T\right) =S_{0}\left(
\mathfrak{F}_{0}\left( \mathcal{Q}\right) \right) =S_{0}\left( \mathcal{Q}%
\right) ,
\end{equation}%
cannot occur, because, through (\ref{eq:fullS}), it would imply%
\begin{gather}
S_{NE}\left( \mathcal{Q};T\right) \simeq S_{0}\left( \mathcal{Q}\right)
+\delta S\left( \mathcal{Q};T\right) =S_{0}\left( \mathcal{Q}\right) \\
\Updownarrow  \notag \\
\delta S\left( \mathcal{Q};T\right) \simeq 0,
\end{gather}%
This cannot true, since, again by means of (\ref{eq:fullS}), it would yield
to%
\begin{equation}
TS_{0}\left( \mathcal{Q}\right) =0,
\end{equation}%
which is impossible, because $T>0$ and\footnote{%
Indeed, we have assumed that the extremal limit yields a \textquotedblleft
large\textquotedblright\ black hole, with attractor mechanism (on the other
hand, from (\ref{eq:fullS}), $S_{0}\left( \mathcal{Q}\right) =0\Rightarrow
S_{NE}\left( \mathcal{Q};T\right) =0~\forall T$, and thus the whole
treatment would be meaningless).} $S_{0}\left( \mathcal{Q}\right) >0$ $%
\blacksquare $
\end{description}

\section{\label{stu}An Example: STU black hole}

In this section, we apply the notion of generalized Freudenthal duality
introduced above to deal with an explicit example. Namely, we consider the $%
STU$ black hole with $D0-D2-D4-D6$ brane charges. The entropy of an extremal
black hole with such charge configuration has the following form ($a=1,2,3$)
\cite{Shmakova:1996nz},
\[
S_{0}=\frac{\pi }{3p^{0}}\sqrt{\frac{4}{3}(\Delta _{a}\tilde{x}%
^{a})^{2}-9(p^{0}p\cdot q-2D)^{2}}
\]%
where $q_{0},q_{a},p^{a},p^{0}$ are the $D0,D2,D4$ and $D6$ brane charges
respectively, $p\cdot q=p^{0}q_{0}+p^{a}q_{a}$, with $\tilde{x}^{a}$
denoting the real solution of the algebraic inhomogeneous system $\Delta
_{a}=D_{abc}\tilde{x}^{b}\tilde{x}^{c}=3D_{a}-p^{0}q_{a}$, and $%
D_{a}=D_{abc}\,p^{b}p^{c}$, $D=D_{a}p^{a}$. Considering an \textit{Ansatz}
that $\tilde{x}^{a}=\sqrt{{\frac{3D-p^{0}\,q_{b}p^{b}}{D}}}p^{a}$ \cite{Chattopadhyay:2021vog}, the extremal entropy can be
written as \cite{Mandal:2017ioi},
\begin{equation}
S_{0}={\frac{\pi }{3p^{0}}}\sqrt{{\frac{4}{3}}{\frac{(3D-p^{0}%
\,q_{a}p^{a})^{3}}{D}}-9(p^{0}\,p\cdot q-2D)^{2}}.  \label{extstuen}
\end{equation}

Here we begin with an extremal $D0-D2-D4-D6$ STU black hole with charge
vectors as $\left( q_{0},q_{a}\right) =(11,5,3,-6)$ and $\left(
p^{0},p^{a}\right) =(7,1,2,9)$. The numerical value of the entropy %
\eqref{extstuen} of this black hole is $S_{0}=259.534$. As it is an extremal
black hole, it has an F-dual with the same value of the entropy, whose
charges can be obtained following \eqref{F-S0} and they are given by $\left(
\tilde{q}_{0},\tilde{q}_{a}\right) =(-3.86709,17.0006,7.78631,11.2499)$ and $%
\left( \tilde{p}^{0},\tilde{p}^{a}\right) =(0.266305,1.42794,2.85587,12.8514)
$.

Now we will show numerically that there also exists a unique near-extremal,
generalised F-dual black hole to another near-extremal black hole, but with
different temperatures. Again, we begin with a $D0-D2-D4-D6$ STU black hole
with charge vector $\left( q_{0},q_{a}\right) =(11,5,3,-6)$ and $\left(
p^{0},p^{a}\right) =(7,1,2,9)$, and with temperature $T=10^{-3}$ ($%
{{}^\circ}%
K$). Inserting these values to derive the entropy of the corresponding
near-extremal black hole, we get the entropy $S_{NE}=289.177$. Following the
formul\ae\ discussed in this paper, we find the F-dual charges as
\begin{eqnarray}
\left( \tilde{q}_{0},\tilde{q}_{a}\right)
&=&(-4.02524,17.6959,8.10476,11.71); \\
\left( p^{0},p^{a}\right)  &=&(0.277196,1.48634,2.97267,13.377),
\end{eqnarray}%
and the F-dual temperature becomes $T+\delta T=0.000238713$ ($%
{{}^\circ}%
K$). Using these charges and temperature, we find the near-extremal entropy $%
\tilde{S}_{NE}=289.177$, which is the same as that of the different
near-extremal black hole, which we have started with.

\section{\label{Conclusion}Conclusion}

In this paper, we have introduced the notion of Freudenthal duality for
\textit{near-extremal} black holes, for which the formula (\ref{eq:newnonen}%
), obtained in \cite{Chattopadhyay:2021vog}, holds true. We have explicitly
proved that in any four-dimensional Maxwell-Einstein theory with action (\ref%
{eq:iiaction}) (corresponding to the purely bosonic sector of any $\mathcal{N%
}\geqslant 2$-extended, $D=4$ supergravity theory\footnote{%
For $\mathcal{N}=2$, excluding hypermultiplets.}), two near-extremal black
holes with temperatures respectively $T$ and $T+\delta T$, and with charges
related by the Freudenthal duality generated by the near-extremal entropy,
can have the \textit{same} entropy (for a given - \textit{unique!} -
analytical\footnote{%
Even if we did not work it out in a completely explicit way (because it does
not yield an illuminating result), we have established that the analytical
expression pertaining to solution $\delta T_{3}$ in formula (\ref{eq:sols})
is unique, which is physically admissible.} expression of $\delta T$ as a the
function of $T$ and of the extremal entropy of one of the two black holes).
We have called this duality map, which non-linearly transforms both the e.m.
charges and the temperature, \textit{near-extremal Freudenthal duality}.

It should be remarked that, even if in our treatment we have considered
doubly-extremal black holes for simplicity's sake, our analysis and the
resulting Eq. (\ref{exen}) are actually holding true for any
\textquotedblleft large\textquotedblright , asymptotically flat, static,
spherically symmetric, dyonic near-extremal black hole in (not necessarily
supersymmetric) Einstein-Maxwell-scalar ungauged theories, endowed with a
geometry of the scalar manifold such that the e.m. charges under
consideration support an attractor solution at the event horizon (in the
extremal limit $T\rightarrow 0^{+}$). This is due to the fact that our
analysis is strictly confined within the near-horizon region, in which (when
assuming the Attractor Mechanism to occur, up to flat directions; cf.
footnote 2) the scalar fields are fixed in terms of e.m. charges for any
extremal black hole.

Moreover, we would like to stress that {temperature is not a fundamental
variable: it can always be expressed in terms of the conserved charges of
the black hole and the asymptotic values of the scalars. Thus, the
transformation under Freudenthal duality of the temperature }${T}${\ should
be consistent with the transformations of those parameters. In fact},
non-extremal black hole do not generally exhibit the Attractor Mechanism, so
their Bekenstein-Hawking entropy depends on both the conserved charges of
the system (i.e., in the case under consideration, on the e.m. charges) and
on the values of the scalar fields at spacial infinity which, in ungauged
theories, can be freely specified. Within the framework considered in the
present paper, the dependence of temperature on the e.m. charges as well as
on the asymptotic values of scalar fields could be obtained as follows : by
considering the general expression of the non-extremal black hole entropy,
one should expand into the non-extremality parameter, and truncate to the
lowest non-trivial order in such a parameter; hence, the comparison with
Eqs. (\ref{exen}) or (\ref{eq:fullS}) would allow to obtain the explicit
expression of the temperature $T$ in the near-extremal limit in terms of the
e.m. charges and of the asymptotic values of scalars. While we consider this
task of utmost interest, our humble opinion is that its thorough
investigation lies beyond the scope of this paper, in which we have dealt
with non-extremality in an effective way, by resorting to temperature within
the JT framework.

So far, Freudenthal duality had been introduced for extremal black holes
only, in both gauged and ungauged supergravity in four space-time
dimensions. Away from extremality (even when the departure from extremality
is slight, such as in near-extremal black holes), entropy generally is no
more homogeneous of degree two in e.m. charges, and a na\"{\i}ve extension
of Freudenthal duality fails. In order to consistently formulate a
Freudenthal duality map for near-extremal black holes, \textit{we have
introduced a (non-linear) transformation of the temperature}, as well. By
exploiting Descartes' rule of signs as well as Sturm's Theorem, which are
excellent tools for analysing the real roots of an algebraic equation
without solving it, we have precisely shown that for a given set of e.m.
charges (which support an extremal black hole with a non-vanishing area of
the event horizon and thus with a non-vanishing Bekenstein-Hawking entropy),
there indeed exist two near-extremal black holes with two unique, small
temperatures such that they share the \textit{same} entropy while their e.m.
charges are related by the Freudenthal duality generated at a \textit{%
non-vanishing} temperature. Thus, our analysis sheds new light on the
invariance properties of the macroscopic entropy of near-extremal black
holes in four space-time dimensions, providing the first example, as far as
we know, of intrinsically non-linear symmetry of the entropy itself.\medskip

In this paper, we have investigated Freudenthal duality for near-extremal
black holes in \textit{ungauged} supergravity (or, more in general, in
\textit{absence} of a potential for scalar fields and of gauging of the
isometries of the associated non-linear sigma model). It would be
interesting to extend our analysis to near-extremal black holes in \textit{%
gauged} supergravity (or, more in general, in presence of a potential for
scalar fields and of gauging of the isometries of the associated non-linear
sigma model).

Moreover, in the present work, we did not consider the extra, \textit{%
logarithmic} correction to the entropy of near-extremal black holes, which
was partially present in the analysis of \cite{Chattopadhyay:2021vog}, and
which was discussed in its (one-loop) full-fledged form recently in \cite%
{Iliesiu:2022onk}. It would be interesting to analyze the notion of
Freudenthal duality for the \textit{logarithmic corrected} entropy of both
extremal and near-extremal black holes. Also, \textit{rotating} extremal
black holes have a Bekenstein-Hawking entropy which is not invariant under
the na\"{\i}vely defined Freudenthal duality, and it would be of interest to
investigate the possibility of a consistent generalization of such an
intrinsically non-linear map to this class of solutions of the
Maxwell-Einstein equations, as well.

It is here worth remarking that in \cite{Galli:2012ji} an alternative
version of Freudenthal duality was put forward, relying on the crucial
observation that the representation of black hole solutions in terms of the $%
H$-variables (which are harmonic functions in the supersymmetric case) is
non-unique, due to the existence of a local symmetry in the effective
action. In \cite{Galli:2012ji} this symmetry is considered as a continuous
(and local) generalization of the Freudenthal duality, which allows to
rewrite the physical fields of a solution in terms of entirely
different-looking functions. While we agree that the near-horizon limit of
the treament of \cite{Galli:2012ji} for near-extremal black holes would
yield to results consistent with the ones obtained within our investigation,
we feel that a thorough study of such a relation would deserve a separate
study, which we leave for future work.

\bigskip
\newpage
\noindent \textbf{Acknowledgments}

The work of AC is supported in part by the South African Research Chairs
Initiative of the National Research Foundation, grant number 78554 and by
the European Union's Horizon 2020 research and innovation programme under
the Marie Sklodowska Curie grant agreement number 101034383.~\ The work of
TM is supported by a Simons Foundation Grant Award ID 509116 and by the
South African Research Chairs initiative of the Department of Science and
Technology and the National Research Foundation. The work of AM is supported
by a \textquotedblleft Maria Zambrano\textquotedblright\ distinguished
researcher fellowship at the University of Murcia, Spain, financed by the
European Union within the NextGenerationEU programme.

\appendix

\section{\label{app:descarte}Descartes' rule of signs}

\textit{Descartes' rule of signs} states that, for a univariate polynomial
function $f(x)$ and the corresponding equation%
\begin{equation}
f(x)=0,  \label{Eqq}
\end{equation}

\begin{itemize}
\item the number of positive real roots of Eq. (\ref{Eqq}) is the same as
(or less than by an even number) the number of changes in the sign of the
coefficients of $f(x)$;

\item the number of negative real roots of Eq. (\ref{Eqq}) is the same as
(or less than by an even number) the number of changes in the sign of the
coefficients of $f(-x)$.
\end{itemize}

Such a rule can be applied to the various cases within the present paper,
namely:

\begin{enumerate}
\item
\begin{equation}  \label{eq:ncubiceq}
f(x)=x^{3}+2x^{2}+2x+\frac{8}{9}.
\end{equation}
Eq. (\ref{eq:ncubiceq}), which is Eq. (\ref{eq:cubiceq}), admits no real
positive roots. Since $f(-x)=-x^{3}+2x^{2}-2x+\frac{8}{9}$, Eq. (\ref%
{eq:cubiceq}) admits three or one real negative root(s).

\item
\begin{equation}  \label{4theq}
f(x)=x^{4}+a_{3}x^{3}+a_{2}x^{2}+a_{1}x+a_{0},
\end{equation}%
with $a_{3},a_{2},a_{1},a_{0}\in \mathbb{R}_{0}^{+}$ . Eq. (\ref{4theq}),
which is Eq. (\ref{eq:quartic}) with $x\equiv \delta T$ and all the
coefficients positive and defined by (\ref{eq:alla}), admits no real
positive roots. Since $f(-x)=x^{4}-a_{3}x^{3}+a_{2}x^{2}-a_{1}x+a_{0}$, Eq. (%
\ref{eq:quartic}) admits four, two or zero real negative roots.

\item
\begin{equation}  \label{new3rd}
f(x)=x^{3}-a_{2}x^{2}+\left( a_{1}a_{3}-4a_{0}\right) x+\left(
4a_{0}a_{2}-a_{1}^{2}-a_{0}a_{3}^{2}\right) ,
\end{equation}
again with $a_{3},a_{2},a_{1},a_{0}\in \mathbb{R}_{0}^{+}$. From (\ref%
{eq:alla}), since $a_{2}>0$, $a_{1}a_{3}-4a_{0}>0$ and $%
4a_{0}a_{2}-a_{1}^{2}-a_{0}a_{3}^{2}<0$, Eq. (\ref{new3rd}), which is Eq. (%
\ref{eq:yeqn}) with $x\equiv y$ and the coefficients defined by (\ref%
{eq:alla}), admits three or one real positive root(s). Since $%
f(-x)=-x^{3}-a_{2}x^{2}-\left( a_{1}a_{3}-4a_{0}\right) x+\left(
4a_{0}a_{2}-a_{1}^{2}-a_{0}a_{3}^{2}\right) $, Eq. (\ref{eq:yeqn}) admits
zero real negative roots.

\item
\begin{equation}  \label{newnew3rd}
f(x)=x^{3}+{\frac{1}{16\pi S_{0}}}x^{2}+{\frac{\left( 37+128\sqrt{\pi S_{0}}%
T\right) }{6912\pi ^{2}S_{0}^{2}}}x-{\frac{25}{2985984\pi ^{3}S_{0}^{3}}}.
\end{equation}%
Eq. (\ref{newnew3rd}), which is Eq. (\ref{eq:y'eqn}) with $x\equiv y^{\prime
}$, admits one real positive root. Since $f(-x)=-x^{3}+{\frac{1}{16\pi S_{0}}%
}x^{2}-{\frac{\left( 37+128\sqrt{\pi S_{0}}T\right) }{6912\pi ^{2}S_{0}^{2}}}%
x-{\frac{25}{2985984\pi ^{3}S_{0}^{3}}}$, Eq. (\ref{eq:y'eqn}) admits two or
zero real negative roots.
\end{enumerate}

\section{\label{app:sturm}Sturm's Theorem}

Sturm's Theorem provides an \textit{algorithmic} way of calculating the
number of simple roots of a non-zero polynomial
\begin{equation}
p(x):=a_{n}x^{n}+a_{n-1}x^{n-1}+\cdots +a_{1}x+a_{0}
\end{equation}%
of degree $n\geqslant 0$ with real coefficients. Before stating Sturm's
Theorem \cite{dorrie1965100,raghavan,Padraic}, let us introduce the following

\paragraph{Definition}

The \textit{canonical sequence} associated to a non-zero polynomial $p(x)$
is defined as the set of polynomials starting from $p(x)$, with the
properties that

\begin{enumerate}
\item If $p_{0}(x)=p(x)$ is a constant polynomial, then the sequence stops
there.

\item $p_{1}(x)=p^{\prime }(x)$.

\item for $i\geqslant 2$, $p_{i}(x)=-$Rem$(p_{i-2}(x),p_{i-1}(x))$ iff Rem$%
(p_{i-2}(x),p_{i-1}(x))\neq 0$. \newline
For non-zero polynomials $a(x)$ and $b(x)$, we denote by Rem$(a(x),b(x))$
the remainder of the Euclidean division of $a(x)$ by $b(x)$.

\item If Rem$(p_{i-2}(x),p_{i-1}(x))=0$, then $p_{i}(x)$ remains undefined,
and the sequence stops there.
\end{enumerate}

There are several consequences of the above definition, but one should note
that canonical sequence starting with $p_{0}(x)$ always stops in less than $%
n $ steps, where $n$ is the degree of $p_{0}(x)$; if the last term in the
sequence is $p_{m}(x)$, with of course $m\leqslant n$, then $p_{m}(x)$ is
equal to the greatest common divisor (gcd) of $p(x)$ and $p^{\prime }(x)$,
up to sign.

\paragraph{Definition}

Let $p_{0}(x),p_{1}(x),\cdots ,p_{m}(x)$ be a non-empty finite sequence of
polynomials, with $p_{0}(x)$ not identically zero. Such a sequence is called
a \textbf{Sturm sequence} iff

\begin{enumerate}
\item The last term $p_m(x)$ of the sequence is either always positive or
always negative on the real line.

\item No two consecutive $p_i(x)$ are simultaneously zero for $x$ a real
number.

\item Suppose that $\alpha$ is a root of $p_i(x)$, for some $i$ with $0<i<m$%
. Then $p_{i-1}(\alpha)$ and $p_{i+1}(\alpha)$ have opposite signs.

\item At any real root $\alpha $ of $p_{0}(x)$, the values of $p_{0}(x)$ at $%
\alpha +0$ and $\alpha -0$ is of opposite sign. This last condition ensures
that $\alpha $ cannot be a repeated root of $p_{0}(x)$.
\end{enumerate}

One can prove that \textit{the canonical sequence associated to a polynomial
without repeated real roots is a Sturm sequence}.

The final ingredient that we need to state Sturm's Theorem is the \textit{%
sign change number function}. Given $\alpha \in \mathbb{R}$ and a polynomial
$p(x)$, the \textbf{number of sign change function} at $\alpha $, denoted by
$\sigma (\alpha )$, as the number of sign changes (ignoring the zeros) in
any Sturm sequence associated to $p(x)$, computed at $x=\alpha $.

We can now state

\paragraph{Sturm's Theorem}

\textit{Let }$p(x)$\textit{\ be a non-zero polynomial with real
coefficients. The number of distinct real roots (counted without
multiplicity) of }$p(x)$\textit{\ in an interval }$(a,b]$\textit{\ of the
real line is given by }$\sigma (a)-\sigma (b)$\textit{, the difference in
the sign change number function at the end points }$a$\textit{\ and }$b$%
\textit{, with respect to any Sturm sequence associated to the given
polynomial }$p(x)$\textit{.\newline
}

When we are interested in looking for real roots of a polynomial $p(x)$
(without repeated real roots) in a given interval of the real line, we can
construct the canonical sequence associated to $p(x)$ (which will be a Sturm
sequence associated to $p(x)$ itself), and calculate the difference between
the values of the sign change number function of $p(x)$ evaluated at the
extrema of the interval, in order to determine the number of real roots of $%
p(x)$ in such an interval. On the other hand, Descartes' rule of signs
applied to $p(x)$ will let us know about the number of roots being on the
positive or negative part of the real axis.

\subsubsection*{Example I}

Let us consider Eq. (\ref{eq:cubiceq}). The canonical sequence associated to%
\begin{equation}
p_{0}(x):=x^{3}+2x^{2}+2x+\frac{8}{9},
\end{equation}%
which in this case is a Sturm sequence associated to $p_{0}(x)$ itself, is
given by
\begin{equation}
\begin{split}
p_{0}(x)& =x^{3}+2x^{2}+2x+{\frac{8}{9};} \\
p_{1}(x)& =2+4x+3x^{2}; \\
p_{2}(x)& =-{\frac{4}{9}}(1+x); \\
p_{3}(x)& =-1.
\end{split}%
\end{equation}%
We want to calculate the number of real roots of the polynomial $p_{0}(x)$,
namely the number of real roots in the interval $(-\infty ,\infty )$.
Therefore we check the sign changes for this sequence at $-\infty $ and $%
\infty $. We find that at $x=-\infty $,

\begin{table}[h!]
\begin{center}
\label{tab:table1}
\begin{tabular}{c|c|c|c}
$p_0$ & $p_1$ & $p_2$ & $p_3$ \\ \hline
- & + & + & - \\
&  &  &
\end{tabular}%
\end{center}
\end{table}
Thus, $\sigma(-\infty)=2$. Similarly at $x=\infty$, we find

\begin{table}[h]
\begin{center}
\label{tab:table2}
\begin{tabular}{c|c|c|c}
$p_0$ & $p_1$ & $p_2$ & $p_3$ \\ \hline
+ & + & - & - \\
&  &  &
\end{tabular}%
\end{center}
\end{table}
implying $\sigma (\infty )=1$. This yields

\begin{equation}
\sigma (-\infty )-\sigma (\infty )=2-1=1,
\end{equation}%
signalling the existence of a single real root between $(-\infty ,\infty )$.
Either using Descartes' rule of signs, or by splitting the interval $%
(-\infty ,\infty )$ in positive and negative semilines and applying Sturm's
Theorem twice, one can finally prove that Eq. (\ref{eq:cubiceq}) admits only
one real root, which is negative.

\subsubsection*{Example II}

Let us consider Eq. (\ref{eq:yeqn}). By considering $a_{0},a_{1},a_{2},a_{3}$
$\in \mathbb{R}_{0}^{+}$ and $B\in \mathbb{R}_{0}^{-}$, the canonical
sequence associated to%
\begin{equation}  \label{eq:ex2}
p_{0}\left( y\right) :=y^{3}-a_{2}y^{2}+(a_{1}a_{3}-4a_{0})y+B,
\end{equation}%
which in this case is a Sturm sequence associated to $p_{0}(y)$ itself, is
given by
\begin{eqnarray}
p_{0}(y) &=&y^{3}-a_{2}y^{2}+\left( a_{1}a_{3}-4a_{0}\right) y+B;  \notag \\
p_{1}(y) &=&3y^{2}-2a_{2}y+B_{1};  \notag \\
p_{2}(y) &=&B_{2}y+B_{3};  \notag \\
p_{3}(y) &=&4a_{0}-a_{1}a_{3}+{\frac{1}{B_{2}^{2}}}\left( {\frac{64}{3}}%
a_{0}a_{1}^{2}a_{2}-{3a_{1}^{4}}-{\frac{1024}{27}}a_{0}^{2}a_{2}^{2}+{\frac{2%
}{3}}a_{1}^{3}a_{2}a_{3}-{\frac{64}{27}}a_{0}a_{1}a_{2}^{2}a_{3}\right.
\notag \\
&&-6a_{0}a_{1}^{2}a_{3}^{2}+{\frac{64}{3}}a_{0}^{2}a_{2}a_{3}^{2}-{\frac{1}{%
27}}a_{1}^{2}a_{2}^{2}a_{3}^{2}+{\frac{2}{3}}%
a_{0}a_{1}a_{2}a_{3}^{3}-3a_{0}^{2}a_{3}^{4}-2B_{2}a_{1}^{2}a_{2}  \notag \\
&&\left. +{\frac{64}{9}}B_{2}a_{0}a_{2}^{2}+{\frac{2}{9}}%
B_{2}a_{1}a_{2}^{2}a_{3}-2B_{2}a_{0}a_{2}a_{3}^{2}\right) ,
\label{Sturm-seq}
\end{eqnarray}%
where, by setting%
\begin{equation}
B:=4a_{0}a_{2}-a_{1}^{2}-a_{0}a_{3}^{2},
\end{equation}%
it follows that
\begin{eqnarray}
B_{1} &:&=-4a_{0}+a_{1}a_{3};  \notag \\
B_{2} &:&=\frac{8}{3}a_{0}+\frac{2}{9}a_{2}^{2}-\frac{2}{3}a_{1}a_{3};
\notag \\
B_{3} &:&=a_{1}^{2}-{\frac{32a_{0}a_{2}}{9}}-{\frac{a_{1}a_{2}a_{3}}{9}}%
+a_{0}a_{3}^{2}.
\end{eqnarray}%
We want to calculate the sign sequence at the three points $y=\{-\infty
,0,\infty \}$. By recalling the definitions (\ref{eq:alla}) and the fact
that both $S_{0}$ and $T$ are real positive (with $T$ also small), we find
the signs of $B,B_{1},B_{2}$ and $B_{3}$ are $\{-,+,-,+\}$, $p_{3}(y)<0$ $%
\forall y\in \mathbb{R}$. With these information, we find that the signs of
the Sturm sequence associated to $p_{0}(y)$ evaluated at $y=\{-\infty
,0,\infty \}$ are respectively given by the following quadruplets : $%
\{-,+,+,-\}$, $\{-,+,+,-\}$, $\{+,+,-,-\}$. Thus, we obtain
\begin{equation}
\sigma (-\infty )=2,\quad \sigma (0)=2,\quad \sigma (\infty )=1,
\end{equation}%
which implies that Eq. (\ref{eq:yeqn}) admits only one real root, which is
positive. Therefore, for any positive and real value of $T$ and $S_{0}$, Eq.
(\ref{eq:yeqn}) always admits only one real root, which is positive.

Analogously, one can show that the Sturm sequence associated to Eq. (\ref%
{eq:y'eqn}) has exactly the same values of $\sigma(\alpha)$ as the example (%
\ref{eq:ex2}), at $\alpha=\{-\infty,0,\infty\}$. This implies that Eq. (\ref%
{eq:y'eqn}) has only one real root, which is positive too.

\end{document}